\documentclass[11pt]{article}
\usepackage{graphicx,color}
\usepackage{chicago}
\usepackage{subfigure}
\usepackage{amsmath, amsfonts, amssymb, amstext}

\newtheorem{lemma}{Lemma}

\newtheorem{remark}{Remark}
\newtheorem{proposition}{Proposition}

\definecolor{DarkBlue}{rgb}{0.1,0.1,0.5}
\definecolor{Red}{rgb}{0.9,0.0,0.1}

\usepackage{setspace}

\def\tr{\mbox{tr}}

\title{Geometric kernel smoothing of tensor fields}

\author{Owen Carmichael$^\dag$, Jun Chen$^\S$, Debashis Paul$^{\S}$
and Jie Peng$^{\S}$\footnote{Email: jie@wald.ucdavis.edu}\\
{\small $^\dag$ Departments of Neuroscience and Computer Science, University of California, Davis}\\
{\small $^\S$Department of Statistics, University of California,
Davis}\\}

\date{}

\begin{document}
\maketitle
\begin{abstract}
In this paper, we study a kernel smoothing approach for
denoising a  tensor field. Particularly, both simulation
studies and theoretical analysis are conducted to understand
the effects of the noise structure and the structure of
the tensor field on the performance of different smoothers
arising from using different metrics, viz.,
Euclidean, log-Euclidean and affine invariant metrics.
We also study the Rician noise model and compare two
regression estimators of diffusion tensors based on
raw diffusion weighted imaging data at each voxel.
\end{abstract}


\section{Introduction}

In this paper, we discuss kernel smoothing of tensor fields by
assuming and utilizing  spatial smoothness of the field. The primary
goal is to reduce the noise from the ``observed" tensors by
combining information from spatial neighborhoods. Here tensors refer
to symmetric positive definite matrices.  They appear in many
contexts, such as fluid dynamics, robotics, tensor-based
morphometry, etc. See Cammoun \textit{et al.} (2009)
for a review on applications of tensors. One of the most important
applications of tensors is in medical imaging. In {\it diffusion weighted
magnetic resonance imaging (DT-MRI)}, diffusion of water molecules is
measured along a pre-specified set of gradient directions (cf. Mori,
2007). The mobility of water molecules in neuronal tissues shows an
anisotropy in the diffusion (cf. Le Bihan et al., 2001). In white
matter, this diffusion anisotropy originates from the faster
diffusion of water molecules through white matter fiber tracts.
Therefore, white matter fiber tracts can by mapped in a non-invasive
way through DT-MRI by revealing the pattern of diffusion. Indeed,
DT-MRI has emerged as among the most important tools in
understanding the structure of the brain. If the diffusion of water
molecules is modeled by a 3-dimensional Brownian motion, then the
covariance matrices of this Brownian motion characterize the
diffusion at the corresponding locations of the brain and thus they
are called {\it diffusion tensors} (see Section \ref{sec:reg_DWI} for more
details on a model for diffusion weighted imaging data).
DT-MRI data are often subject to a substantial amount of noise,
which results in a noisy tensor field (Gudbjartsson and Patz, 1995;
Hahn \textit{et al.}, 2006, 2009; Zhu \textit{et al.}, 2007, 2009).
This may adversely affect the mapping of white matter fiber tracts
(the process is known as {\it tractography}) and other applications
relying on diffusion tensors as input data, such as measuring the
anisotropy, morphometry etc. (Basser and Pajevic, 2000;  Zhu
\textit{et al.}, 2009). On the other hand, the physical continuity
of the white matter implies a certain degree of spatial smoothness
of the tensor field, which can be utilized to denoise the data.

One of the most commonly used techniques for denoising data
measured at different spatial locations (or different time points)
is through weighted averages over spatial (or temporal)
neighborhoods. The weights are often determined by a
kernel function and thus this approach is referred to
as {\it kernel smoothing}. In this paper, we extend kernel
smoothing techniques to the tensor space. Since the concept
of averaging depends on the geometry of the space, we study the
impacts of geometry on tensor smoothing. In particular,
we consider three geometries on the tensor space: Euclidean
geometry, log-Euclidean geometry and the affine invariant geometry.
Under the Euclidean geometry, the average of the tensors is simply
the (ordinary) arithmetic mean. The log-Euclidean geometry on a tensor
space is proposed by Arsigny \textit{et al.} (2005, 2006), where the
averaging amounts to taking Euclidean average of the logarithm
of the tensors followed by exponentiation of the resulting matrix.
The affine-invariant geometry corresponds to the canonical
Riemannian metric on a tensor space treated as a homogeneous space (Nomizu, 1954).
In context of DT-MRI, it has been studied  by F\"{o}rstner and Moonen
(1999), Fletcher and Joshi (2004, 2007), Pennec \textit{et al.} (2006).
There is no closed form formula for the affine invariant  average
(see Section \ref{sec:kernel_smoothing} for more details). In  Arsigny et al. (2005, 2006), the
authors compare these three geometries when used in various procedures
applied to the tensor field such as interpolation and filtering. It
is shown there that, despite its simplicity,  the Euclidean geometry
has several drawbacks, in particular the swelling effect,
which refers to the phenomenon of the interpolated tensor having a larger
determinant than those of both the original tensors. In the context
of DT-MRI, the determinant of a tensor reflects the degree of
dispersion of water molecules, and thus the swelling effect is
contradictory to physical principles. On the contrary, both the
log-Euclidean and the affine-invariant geometries do not suffer from this
problem. Indeed, under both geometries, tensor interpolation
leads to the interpolation of the determinants (Arsigny \textit{et al.},
2005). Moreover, Arsigny \textit{et al.} (2005, 2006) show that,
the log-Euclidean and the affine-invariant geometries give similar
results for various tensor computations, with the former being
computationally easier.

Along with spatial smoothness, tensor fields often show certain
degrees of anisotropy, e.g., due to faster water diffusion along
white matter fiber tracts in context of DT-MRI. Conventional kernel
smoothing methods use spherical neighborhoods and perform same
amount of smoothing along every direction (referred to as the
{\it isotropic smoothing}). On the contrary, {\it anisotropic  smoothing}
methods impose relatively larger amount of smoothing along the
directions that show a greater degree of anisotropy (such as along
the fiber tracts). Thus such methods better preserve structures of the tensor field
 compared to isotropic smoothing. This
is particularly desirable for DT-MRI since smoothed tensors and
fractional anisotropies computed from them are often used as inputs
for tractorgraphy.  Anisotropic smoothing has been considered in
Chung \textit{et al.} (2003, 2005), Tabelow \textit{et al.} (2008).

In this paper, we conduct both simulation studies
(Section \ref{sec:simulation}) and theoretical analysis
(Section \ref{sec:perturbation_smoothing}) to explore the
impacts of geometry on tensor smoothing.
We find that, the (relative) performances of smoothing methods under different geometries mainly depend on two factors: (i) the noise
structure; (ii) the structure of the tensor field. With additive or
approximately additive noise, Euclidean smoothing perform comparably
(under small noise levels) or better than (under moderate to large noise levels)
the log-Euclidean smoothing and the affine-invariant smoothing
(hereafter, referred to as {\it geometric smoothing}). This is partly due to the fact that
both geometric smoothing methods involve taking logarithm of the
tensors which amplifies additive noise. Here, ``additive noise"
refers to the noise structure where the expectation of the noise-corrupted
tensor equals to the underlying noiseless tensor (see Remark \ref{rem:additive_noise} in
Section \ref{sec:perturbation_smoothing}).  On the other hand, when the
noise structure is highly non-additive, geometric smoothing tend to outperform
Euclidean smoothing. In terms of  tensor field structures, Euclidean smoothing works
comparably to geometric smoothing in homogeneous regions,
especially those with predominantly (nearly) isotropic tensors, while the geometric
smoothing works better in regions with relatively high degrees
of heterogeneity. The perturbation analysis conducted in Section \ref{sec:perturbation_smoothing}
 provides theoretical justifications for some findings of the simulation studies.
Apart from highlighting the difference between the
Euclidean and geometric smoothing, the perturbation analysis results also indicate
that when the tensors are nearly isotropic, the two geometric smoothing
methods are similar, which is also pointed out by Arsigny \textit{et
al.} (2005, 2006).
The simulation results also point to  the need
of a multi-scale approach, i.e., using adaptively chosen
bandwidths for different regions according to their
degrees of homogeneity of the tensors.  For example,
in the presence of crossing fibers, or near the edge of the fiber
bundles (which are regions of high degrees of heterogeneity), tensor smoothing,
especially  under the Euclidean geometry, could lead to even  worse
results compared to the (un-smoothed) observed noisy tensors unless very small bandwidths are used. Whereas, in highly homogenous regions, relatively larger bandwidths often lead to better results.
A multi-scale approach for tensor smoothing is proposed and studied in
Polzehl and Spokoiny (2006) and Tabelow
\textit{et al.} (2008). Finally, the simulation studies show that  in presence of highly
anisotropic tensors, anisotropic smoothing usually improves over
isotropic smoothing.

In this paper, we also conduct numerical (Section \ref{sec:simulation}) and theoretical (Section \ref{sec:reg_DWI}) studies
under the Rician noise model to compare linear
and nonlinear regression methods for extracting tensors from
gradient-based diffusion weighted imaging data.
These studies show that, the nonlinear regression method improves over the linear
regression method, since the former has lower variability when the signal-to-noise
ratio (SNR) is large. Indeed, when SNR is large ,
the efficiency of the nonlinear regression estimator is comparable to that of the
maximum likelihood estimator. Zhu \textit{et al.} (2009) also carry out
asymptotic analysis of regression estimators under the Rician noise model, although it is under a different asymptotic framework.

The rest of the paper is organized as follows. In Section 2, we discuss kernel smoothing on a tensor space. In Section 3, we present results from simulation studies and discuss their implications. In Section 4, a perturbation analysis is conducted to compare the means of a set of tensors under three different geometries on the tensor space. In Section 5, asymptotic analysis is conducted to study and compare the regression estimators under the Rician noise model. Technical details are given in the appendix.

\section{Kernel Smoothing on Tensor
Space}\label{sec:kernel_smoothing}

In this section, we extend kernel smoothing to a smooth
manifold. In particular, we consider smoothing on the space of $N
\times N$ positive definite matrices (hereafter, referred to as the
tensor space, and denoted by $\mathcal{P}_N$). Consider a function
$f:  \mathcal{D} \subset \mathbb{R}^d \rightarrow  \mathbb{R}^p$. Suppose that we observe
pairs $\{(s_i, X_i)\}_{i=1}^n$ with $s_i \in {\cal D}$ being the
design points, and $X_i \in \mathbb{R}^p$ being noise-corrupted
versions of $f(s_i)$. The goal is to reconstruct the function $f$
based on such observations. One way to fit $f(\cdot)$ at location $s
\in  \mathcal{D}$ is to locally approximate $f(\cdot)$ by a
polynomial and then use suitably weighted data
to determine the coefficients of this polynomial. In statistics,
this is called {\it local polynomial smoothing} (Fan and Gijbel, 1996).
The simplest form is to consider a constant function approximation.
Specifically, for $s \in \mathcal{D}$ ,
\begin{eqnarray}\label{eq:kernel_smooth_ave}
\widehat{f}(s):={\rm arg}\min_{c} \sum_{i=1}^{n} \omega_i(s)
\parallel X_i-c
\parallel_2^2,
\end{eqnarray}
where $\parallel \cdot \parallel_2$ denotes a Euclidean norm, and
$\omega_i(s) \geq 0$ are nonnegative weights.  A common scheme for
the weights is
\begin{equation}
\label{eq:isotropic} \omega_i(s):=K\left(\frac{s_i-s}{h}\right), ~~~i=1,\cdots,n,
\end{equation}
where $K(\cdot)$ is a nonnegative, smooth kernel on $\mathbb{R}^d$,
and $h > 0$ is the bandwidth. Thus this method is also called kernel
smoothing. Note that, $\parallel X_i-c
\parallel_2$, denoted by $d(X_i,c)$, is simply the Euclidean distance between $X_i$ and
$c$. Let $(\mathcal{M},g)$ denote  a  smooth manifold $\mathcal{M}$
equipped with a metric $g$ (i.e., a Riemannian manifold). Also
denote the corresponding geodesic distance  by
$d_{\mathcal{M}}(\cdot,\cdot)$. Then kernel smoothing can be
immediately generalized to $(\mathcal{M},g)$ by using $d_{\mathcal{M}}(\cdot,\cdot)$ in place
of $d(\cdot,\cdot)$ in definition (\ref{eq:kernel_smooth_ave}).
Specifically, for a function $f: \mathcal{D} \subset \mathbb{R}^d \rightarrow
(\mathcal{M},g)$, and the observed data pairs $\{(s_i,
X_i)\}_{i=1}^n$ where $s_i \in \mathcal{D}$ and $X_i \in
\mathcal{M}$ are noise-corrupted versions of $f(s_i)$,
the kernel smoothing of $f(\cdot)$ is
\begin{eqnarray}\label{eq:Karcher_local_const}
\widehat{f}(s):={\rm arg}\min_{Y \in \mathcal{M}} \sum_{i=1}^{n}
\omega_i(s) d^2_{\mathcal{M}}(X_i,Y), ~~~ s \in \mathcal{D},
\end{eqnarray}
which is in the form of a weighted Karcher mean
(Karcher, 1977) of $\{X_1,\cdots, X_n\}$. Note that, different
geometries may be imposed onto a smooth manifold $\mathcal{M}$,
which in turn result in different geodesic distances on $\mathcal{M}$. Since
the set of $N \times N$ positive definite matrices is the interior
of a cone (consisting of all positive semi-definite $N \times N$ matrices) in the
Euclidean space $\mathbb{R}^{N \times N}$, Euclidean geometries can
be naturally imposed onto the tensor space and the corresponding
distances are referred to as the Euclidean distances.  One such
example is $d_E(X,Y):=\{{\rm tr}(X-Y)^2\}^{1/2}$ which corresponds
to the trace norm on $\mathbb{R}^{N \times N}$. Under Euclidean
distances, (\ref{eq:Karcher_local_const}) can be easily solved by a
weighted average
\begin{eqnarray}
\label{eq:local_constant_Euclidean}
\widehat{f}_E(s)=\sum_{i=1}^{n} \omega_i(s)
X_i\Bigl/\sum_{i=1}^{n}\omega_i(s),
\end{eqnarray}
where the addition and scalar multiplication are the usual matrix
addition, and scalar multiplication, respectively. As an alternative
to Euclidean geometries, Arsigny \textit{et al.} (2005, 2006) propose
logarithmic Euclidean (henceforth {\it log-Euclidean}) geometries on the
tensor space. Firstly, a Lie group structure is  added to the tensor
space through defining a logarithmic multiplication $\odot$
$$
X \odot Y :=\exp\left(\log (X) + \log (Y)\right), ~~ X, Y \in \mathcal{P}_N.
$$
Then it is shown that, the bi-invariant metrics on
this tensor Lie group exist and lead to distances of the form:
\begin{equation}\label{eq:logE_metric}
d_{LE}(X, Y) = \parallel \log X -\log Y \parallel,
\end{equation}
where $\parallel \cdot \parallel$ is a Euclidean norm. Note that,
the distance defined in (\ref{eq:logE_metric}) is similarity-invariant if the trace norm is
used.  Under these log-Euclidean distances,
(\ref{eq:Karcher_local_const}) can also be explicitly solved as
\begin{equation}\label{eq:local_constant_log_Euclidean}
\widehat{f}_{LE}(s) = \exp\left(\sum_{i=1}^{n} \omega_i(s) \log
(X_i)\Bigl/\sum_{i=1}^{n}\omega_i(s)\right),
\end{equation}
i.e., the matrix exponential of the weighted average of the
logarithm of the tensors.

It is also well-known that $\mathcal{P}_N$ can be identified
with the quotient space $GL^{+}(N,\mathbb{R})/SO(N,\mathbb{R})$ which is
defined by the conjugate group action $\theta$ of $ GL^{+}(N,\mathbb{R})$ on
$\mathcal{P}_N$:
\begin{equation*}
\theta(g, X):=g X g^{T}, ~~~\mbox{for}~~~ g \in GL^{+}(N,\mathbb{R})
~~\mbox{and}~~ X \in \mathcal{P}_N.
\end{equation*}
Here, $GL^{+}(N,\mathbb{R})$ is the identity component of the general linear
group $GL(N,\mathbb{R})$ --  the Lie (sub)group consisting of
$N \times N$ matrices with positive determinant; and $SO(N,\mathbb{R})$
is the special orthogonal group -- the Lie (sub)group consisting of
$N \times N$ orthogonal matrices with determinant
one.  The above quotient space (also referred to as $\mathcal{P}_N$)
is a \textit{naturally reductive homogenous
space} (Absil \textit{et al.}, 2008) and its bi-invariant
metric is given by: for $X \in \mathcal{P}_N$ and $S,
T \in  T_X(\mathcal{P}_N)$ -- the tangent space of $ \mathcal{P}_N$ at $X$,
\begin{equation*}
\langle S,T \rangle_X :={\rm tr}\left(S X^{-1} T X^{-1}\right).
\end{equation*}
Under this metric, the geodesic distance between two points $X, Y
\in \mathcal{P}_N$ is
\begin{equation}\label{eq: affine_metric}
d_{Aff}(X, Y) :=\left[\tr\left(\log (X^{-1/2} Y
X^{-1/2})\right)^2\right]^{1/2} = \left(\langle  \log_X(Y),
\log_X(Y) \rangle_X\right)^{1/2},
\end{equation}
where $\log_X(Y) := X^{1/2} \log (X^{-1/2} Y X^{-1/2}) X^{1/2}$
is the logarithm map on this space.
Since this metric is affine invariant, i.e., for any $g \in
GL^{+}(N,\mathbb{R})$
$$
d_{Aff}(\theta(g,X), \theta(g,Y)) = d_{Aff}(X, Y),
$$
hereafter, it is referred to as the {\it affine-invariant metric}.

The affine-invariant geometry on $\mathcal{P}_N$ has been
extensively studied. For example,  by F\"{o}rstner and Moonen
(1999), Fletcher and Joshi (2004, 2007).
Particularly, in Pennec \textit{et al.} (2006), a Riemannian
framework for tensor computations is proposed under the
affine invariant geometry and is applied to DT-MRI studies. Since
$\mathcal{P}_N$ has non-positive curvature (Skovgaard, 1984), as
pointed out by Pennec \textit{et al.} (2006), there exists a unique
solution to (\ref{eq:Karcher_local_const}) under the
affine-invariant metric (\ref{eq: affine_metric})  if all weights
are positive. An intrinsic gradient descent algorithm is proposed in
Pennec \textit{et al.} (2006)  for solving the weighted Karcher mean
problem on this space. Alternatively, in Ferreira \textit{et al.}
(2006), and Fletcher and Joshi (2007), intrinsic Newton-Raphson
algorithms are derived to find the Karcher mean, which can also be
easily adapted to solve (\ref{eq:Karcher_local_const}). In this
paper, we propose a simple iterative method which does not involve
any (intrinsic) gradient or Hessian computations. This method is
based on the following observation: for a sequence of points
$z_1,\ldots,z_n$ in the Euclidean space $\mathbb{R}^p$, one can
compute their weighted mean $m := (\sum_{i=1}^n w_i z_i)/
(\sum_{i=1}^n w_i)$ for $w_1,\ldots,w_n \geq 0$ through an $n$-step
recursive procedure:
\begin{itemize}
\item[(i)] Set $m_1 = z_1$, and $j=1$;
\item[(ii)] Compute $m_{j+1} = m_j + \frac{w_{j+1}}{\sum_{i=1}^{j+1} w_i}
(z_{j+1} - m_j)$;
\item[(iii)] If $j+1 < n$, set $j=j+1$ and go to (ii). Otherwise, set $m = m_n$ and stop.
\end{itemize}
In step (ii), $m_{j+1}$ is on the line segment connecting $m_j$ and
$z_{j+1}$. Note that straight lines are geodesics in the Euclidean
space. Thus, as a generalization to a Riemannian manifold $(\mathcal{M},g)$,
this step can by modified  as follows:
\begin{itemize}
\item[$(ii)^{'}$] Compute
$m_{j+1}=\gamma(1-\frac{w_{j+1}}{\sum_{i=1}^{j+1} w_i})$, where
$\gamma(\cdot)$ is the geodesic on $(\mathcal{M},g)$ such that
$\gamma(0)=m_j$ and $\gamma(1)=z_{j+1}$.
\end{itemize}
Note that, in general, the mean value computed this way depends on
the ordering of the points $z_1,\ldots,z_n$ unless  $z_i$'s all lie
on a geodesic in $(\mathcal{M},g)$. When solving
(\ref{eq:Karcher_local_const}), we propose to pre-order $X_i$'s
based on the Euclidean distances of $s_i$'s from $s$, i.e.,
$z_1:=X_1$, where $s_1$ is closest to $s$ among all $s_i$'s, etc.
For a manifold $(\mathcal{M},g)$, the geodesic $\gamma(\cdot)$ between
two (nearby) points $X, Y$ such that $\gamma(0)=X$ and $\gamma(1)=Y$
can be expressed through the exponential map and the logarithm map:
$\gamma(t)= \exp_X(t\log_X(Y))$. For the tensor space with
affine-invariant metric, we have: for $X, Y  \in \mathcal{P}_N$ and
$S \in T_X(\mathcal{P}_N)$
\begin{equation}\label{eq:exp_log_affine_invariant}
\exp_X(S)=X^{1/2}\exp(X^{-1/2}S X^{-1/2})X^{1/2}, ~~~
\log_X(Y)=X^{1/2} \log(X^{-1/2}Y X^{-1/2}) X^{1/2}.
\end{equation}
Compared with the gradient-based methods, the above algorithm has the
advantages of being numerically stable and computationally
efficient. In practice, it leads to similar results when solving
(\ref{eq:Karcher_local_const}) as those obtained by the
gradient-based methods (results not shown).

In this paper, we refer to the smoothing methods using the metrics
defined in (\ref{eq:logE_metric}) or (\ref{eq: affine_metric}) as
{\it geometric smoothing}, whereas the smoothing methods using the
Euclidean metrics as {\it Euclidean smoothing}. The two geometric
metrics have some appealing properties such as invariance.
Moreover, under both log-Euclidean metric and
affine-invariant metric, interpolation of tensors results in an
interpolation of their determinants (Arsigny \textit{et al.}, 2005).
Thus, unlike the Euclidean smoothing, geometric smoothing methods do
not suffer from swelling effects. However, as we shall
see in Sections \ref{sec:simulation} and \ref{sec:perturbation_smoothing}, the
relative merits of these metrics in terms of tensor smoothing are
less obvious and they rely on several factors (and their
interactions), especially, the noise structure and the structure of
the tensor field being smoothed. It turns out that, Euclidean
smoothing perform comparably or even better than geometric smoothing, under nearly additive noise and/or regions dominated by
(nearly) isotropic tensors. The perturbation analysis performed
in Section \ref{sec:perturbation_smoothing} shows that, the logarithm operation on the
tensors (which is used for both geometric smoothing methods, but not
in Euclidean smoothing) amplifies additive noise, and consequently
results in a bias. On
the other hand, if the noise structure is highly non-additive and/or
the regions are with heterogeneous tensors (such as at the
crossings of fiber bundles  or on the boundary of fiber bundles), geometric smoothing
tend to outperform Euclidean smoothing, presumably as a benefit of
respecting the intrinsic geometry of the tensor space.

Besides the choice of metrics, one also needs to
choose a scheme to assign weights $w_i(s)$'s in
(\ref{eq:Karcher_local_const}). The conventional approach is to
simply set weights as in (\ref{eq:isotropic}) where $K$ is a fixed
kernel, e.g., the Gaussian kernel $k(t)=\exp(-t^2/2)$. This is
referred to as \textit{isotropic smoothing}. However, the tensor
field often shows various degrees of anisotropy in different regions
and the tensors tend to be more homogeneous along the leading
anisotropic directions (e.g., along the fiber tracts). Thus it makes
sense to set the weights larger if $s_i-s$ is along the leading diffusion
direction at $s$. Therefore, we propose the following
\textit{anisotropic weighting scheme}:
\begin{equation}\label{eq:weight_aniso_trace}
\omega_i(s) := K_h(\sqrt{{\rm tr}(\widehat{D}) (s_i-s)^T
\widehat{D}^{-1}(s_i-s)}),
\end{equation}
where $\widehat{D}$ is the current estimate of the tensor at voxel
location $s$, and $K_h(\cdot) := K(\cdot/h)$ is a nonnegative,
integrable kernel with $h>0$ being the bandwidth. The use of ${\rm
tr} (\widehat{D})$ in (\ref{eq:weight_aniso_trace}) is to set the
weights scale-free with regard to $\widehat{D}$. There are other
schemes for anisotropic weights. For example, in
Tabelow \textit{et al.} (2008), the term ${\rm tr}(\widehat{D})$ is
replaced by $\det(\widehat{D})$ in (\ref{eq:weight_aniso_trace}),
which is supposed to capture not only the directionality of the
local tensor field, but also the degree of anisotropy.
Chung \textit{et al.} (2003, 2005) also propose kernel smoothing under
Euclidean geometry with anisotropic kernel weights.
Simulation studies in Section \ref{sec:simulation} show that,
as expected, anisotropic smoothing improves over isotropic
smoothing in regions with highly anisotropic tensors,
and the two perform similarly in regions with predominantly
isotropic tensors.  Moreover, for both isotropic weights
(\ref{eq:isotropic}) and anisotropic weights (\ref{eq:weight_aniso_trace}), the
bandwidth $h$ needs to be specified. The bandwidth $h$ controls the amount of
smoothing: the larger $h$ is, the more smoothing is performed. The
choice of bandwidth is an active field of research in kernel
smoothing. In principle, in relatively more homogeneous regions, a
larger $h$ should be used, while in relatively less homogeneous
regions, a smaller $h$ should be used. Since the tensor fields often show various degrees
of homogeneity  across different regions, i.e., in some regions, the neighboring tensors are
more alike, while in some other regions, the neighboring tensors are more
different, bandwidth should be adaptively chosen according to the local
degree of homogeneity.  This idea has been studied in Polzehl and Spokoiny (2006),
Polzehl and Tabelow (2008) and Tabelow \textit{et
al.} (2008), where a structural adaptive smoothing is proposed.
In this paper, we do not try to
give a general prescription of bandwidth selection for tensor
smoothing. Rather, we simply show by simulation studies that the
smoothing methods work best under different bandwidth in different
regions.

\section{Simulation Studies}\label{sec:simulation}

In this section, we conduct systematic simulation studies to explore
the impacts of geometry on tensor smoothing. We also
compare isotropic and anisotropic smoothing, as well as examine
tensor smoothing across a set of  bandwidth choices. We first
describe the simulation design, in particular, the structure of the
simulated tensor field and the noise models.

\subsection{Simulated tensor field}\label{subsec:simul_design}

\begin{figure}[h]
\begin{center}
\includegraphics[width=3in, height=3in, angle=0]{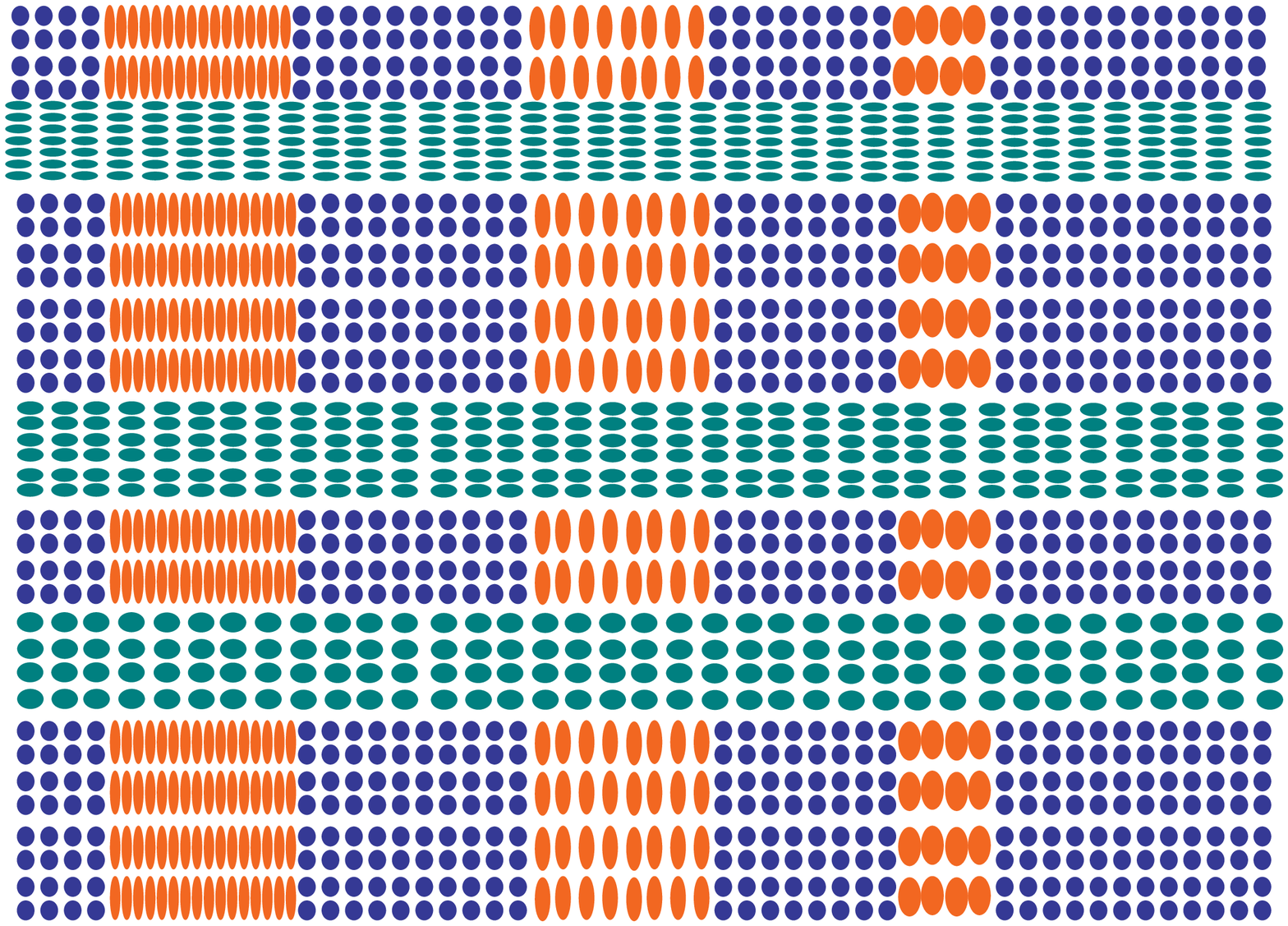}
\caption{Schematic plot of one slice of the simulated tensor field.}
\label{fig:wide_band_design}
\end{center}
\end{figure}

The simulated tensor field consists of $4$ slices (in the
$z$-direction) each of dimension $128 \times 128$ (in the $x$- and
$y$-directions). Thus, altogether there are $128\times 128 \times 4$
voxels. The tensor field consists of  two types of regions: the
{\it background regions} with identical isotropic tensors -- the $3 \times
3$ identity matrices $I_3 =$ diag$(1,1,1)$; and  the {\it bands regions}
with anisotropic tensors: on each band, all the tensors are the same
and they point to either the vertical (i.e., $y$-direction) or the
horizontal direction (i.e., $x$-direction).  More specifically, each
slice has three parallel vertical bands with tensors all point to
the vertical direction and three horizonal bands with tensors all
point to the horizontal direction.  These bands are of various
widths and the tensors on them also show various degrees of
anisotropy. Whenever a horizontal band and a vertical band cross
each other, the tensors at the crossing voxels are set as the ones
on the corresponding horizontal band. See schematic plot Figure
\ref{fig:wide_band_design} for an illustration. Moreover, the bands
structures for slices 1 and 2 are the same, and those of slices 3 and 4
are the same. The location and width of the bands and the tensors on
each band are given in Table 1. 

\begin{table}[h]\label{table:wide_band_design}
\begin{center}
\caption{Description of the simulated tensor field}
\begin{tabular}{c|c|c|c|c}
\hline Band orientation & Slice \# & Band \# & Indices for band boundaries & Tensor on band \\
\hline Horizontal & 1 and 2 & 1 & 20, 35 &
\mbox{diag}$(0.25,16,0.25)$ \\
& & 2 & 60, 75 & \mbox{diag}$(0.5,4,0.5)$ \\
&  & 3 & 90, 105 & \mbox{diag}$(0.7,2,0.7)$ \\
\cline{2-5}
 & 3 and 4 & 1 & 40, 50 &
\mbox{diag}$(0.25,16,0.25)$ \\
& & 2 & 80, 90 & \mbox{diag}$(0.5,4,0.5)$ \\
&  & 3 & 110, 120 & \mbox{diag}$(0.7,2,0.7)$ \\
\hline Vertical & 1 and 2 & 1 & 20, 35 &
\mbox{diag}$(16,0.25,0.25)$ \\
& & 2 & 60, 75 & \mbox{diag}$(4,0.5,0.5)$ \\
& & 3 & 90, 105 & \mbox{diag}$(2,0.7,0.7)$ \\
\cline{2-5}
 & 3 and 4 & 1 & 40, 50 &
\mbox{diag}$(0.25,16,0.25)$ \\
& & 2 & 80, 90  & \mbox{diag}$(0.5,4,0.5)$ \\
& & 3 & 110, 120 & \mbox{diag}$(0.7,2,0.7)$ \\
\hline
\end{tabular}
\end{center}
\end{table}

\subsection{Noise models}\label{subsec:simul_noise}

After the tensors (referred to as the {\it true tensors}) are
simulated according to the previous section, we add noise to them to
derive noise corrupted tensors (referred to as the
{\it observed tensors}). We consider two different noise structures. The
first noise model is motivated by the process of data acquisition in
DT-MRI studies from diffusion weighted imaging data. The second
noise model is motivated by the consideration of decoupling the
noise variability in the eigenvalues and eigenvectors. We describe
both models in detail in the following.



\subsubsection*{Rician noise }

In this model, we first generate the noiseless diffusion-weighted
multi-angular images (DWI) data via model (\ref{eq: dwi_gaussian}) (Mori, 2007)
 using a set of $9$
gradient directions, i.e., for a voxel with true diffusion
tensor $D$, the diffusion weighted signal intensity in direction $b_k$ (a $3\times 1$
vector of unit norm) is modeled as
\begin{equation}
\label{eq: dwi_gaussian}
\overline{S}_k = S_0 \exp\left(- b_k^T D b_k\right),
\end{equation}
where $S_0$, or the baseline signal strength (without diffusion weighting), is set to be $10$ in
our simulation. These $9$ gradient directions come from  a real
DT-MRI study, and they are the normalized versions of the following
vectors:
$$
(1,0,1),~(1,1,0),~(0,1,1),~(0.3,0.2,0.1),~(0.9,0.45,0.2),~(1,0,0),~
(0,1,0),~(0,0,1),~(2,1,1.3).
$$
In our simulation, each of the 9 gradient directions is repeated
twice which results in $18$ DWI measurements per tensor. (This is
the design used in the aforementioned real study). These DWI
intensities $\overline{S}_k$ are then corrupted by Rician noise (see
equation (\ref{eq:S_b_def}) in Section \ref{sec:reg_DWI}), which
gives rise to noise corrupted DWI's  $S_k$'s (referred to as
the {\it observed DWI's}) for each tensor. We consider three different
values for the noise parameter $\sigma$ in equation
(\ref{eq:S_b_def}), namely, $\sigma=0.1, 0.5, 1$ which corresponds to low, moderate and high noise
levels, respectively. Finally, at each voxel, a regression procedure
is applied to the observed DWI data to derive the observed tensor.
Two different regression procedures are considered (i) linear
regression of $\{\log(S_k)\}$ on $\{b_k\}$ (see equation (\ref{eq:log_LS})); (ii) nonlinear
regression of $\{S_k\}$ on $\{b_k\}$ (see equation (\ref{eq:log_NL})). Note that, unlike smoothing methods, for each voxel, regression procedures only use DWI data
from that voxel to derive the corresponding tensor and no information from
neighboring voxels is used. See Section \ref{sec:reg_DWI} for more
details on the regression procedures.


The analysis carried out in Section \ref{sec:reg_DWI} also shows
that, under the Rician noise model, at relatively high
signal-to-noise ratios, the observed DWI intensities as well as the
tensors derived from them by regression methods are approximately unbiased, i.e.
$\mathbb{E}(S_k) \approx \bar{S}_k$, and
$\mathbb{E}(\widehat{D}) \approx D$. A high signal-to-noise ratio
results from either a small noise level for each intensity measurement
(i.e., a small $\sigma$), or a large number of
gradient directions. This implies that, under such cases, we can view Rician noise for tensors  as
an approximately additive noise (See also Remark \ref{rem:Rician_additive} in Section \ref{sec:reg_DWI}).

\subsubsection*{Spectral noise}

In this model, tensors are corrupted directly by adding random
perturbations to their eigenvalues and eigenvectors which leads to a
highly non-additive noise structure. Specifically, consider the
spectral decomposition of the tensor $D$
\begin{equation*}
D = E \Lambda E^T,
\end{equation*}
where $\Lambda$ is a diagonal matrix with positive non-increasing
diagonal elements (i.e., eigenvalues of $D$), and $E$ is the
corresponding orthogonal matrix consisting of eigenvectors of $D$.
Then, the noise corrupted tensor $\widetilde D$ is derived by
\begin{equation*}
\widetilde D := \widetilde E \widetilde \Lambda \widetilde E^T.
\end{equation*}
In the above, $\widetilde \Lambda$ is a diagonal matrix, with
$\widetilde \Lambda_{jj} = \Lambda_{jj} U_j$ ($j=1,2,3$)  where
$U_j$'s are i.i.d. $ \chi_{(\nu)}^2/ \nu$. Here $\nu$ is a
positive integer and $\chi_{(\nu)}^2$ denotes the Chi-square
distribution with $\nu$ degrees of freedom.  The parameter $\nu$
controls the degree of fluctuation of $\widetilde \Lambda$ around
$\Lambda$. Indeed, $\mathbb{E}(\widetilde \Lambda) = \Lambda$, and larger
$\nu$ implies smaller fluctuations. Moreover,
\begin{equation*}
\widetilde E := (I_3+\eta Z) [(I_3+\eta Z)^T(I_3+\eta Z)]^{-1/2}  E
\end{equation*}
where $Z$ is a $3 \times 3$ matrix with i.i.d. $N(0,1)$ entries,
which are independent of $\{U_j\}_{j=1}^3$, and $\eta > 0$. Note
that, $(I_3+\eta Z) [(I_3+\eta Z)^T(I_3+\eta Z)]^{-1/2}$ is a random
orthogonal matrix. The parameter $\eta$ controls the degree of this
matrix deviating  from the identify matrix $I_3$ and  larger $\eta$
implies more fluctuations in $\tilde{E}$. In the simulation, we
consider the following values for the noise parameters $(\nu,\eta)$:
(50,0.1), (50,0.2) (50,0.3), (20,0.1), (20,0.2) and (20,0.3).


\subsection{Results of tensor smoothing}\label{subsec:wide_band_results}

In this section, we report the estimation errors in terms of the
affine-invariant distance (\ref{eq: affine_metric}) between the true
and smoothed tensors at each voxel.  We observe that these errors
across the tensor field have a right-skewed distribution (more skewed for
the geometric smoothers). Thus, measures like mean and standard
deviation are not very representative of the error distribution. Therefore, the median and median
absolute deviation about median (abbreviated as MAD) of the errors
are used as performance measures. Median quantifies
the typical error across tensors and MAD measures the robustness of
the smoothers. Under the Rician noise model, for geometric
smoothing, the MAD of the errors is somewhat larger than that of
Euclidean smoothing, especially at higher noise levels, indicating
that geometric smoothing is more sensitive to nearly additive noise.
The situation is reversed for the spectral noise model.

\begin{figure}[h]
\begin{center}
\includegraphics[width=4.5in, height=3.5in,angle=0]{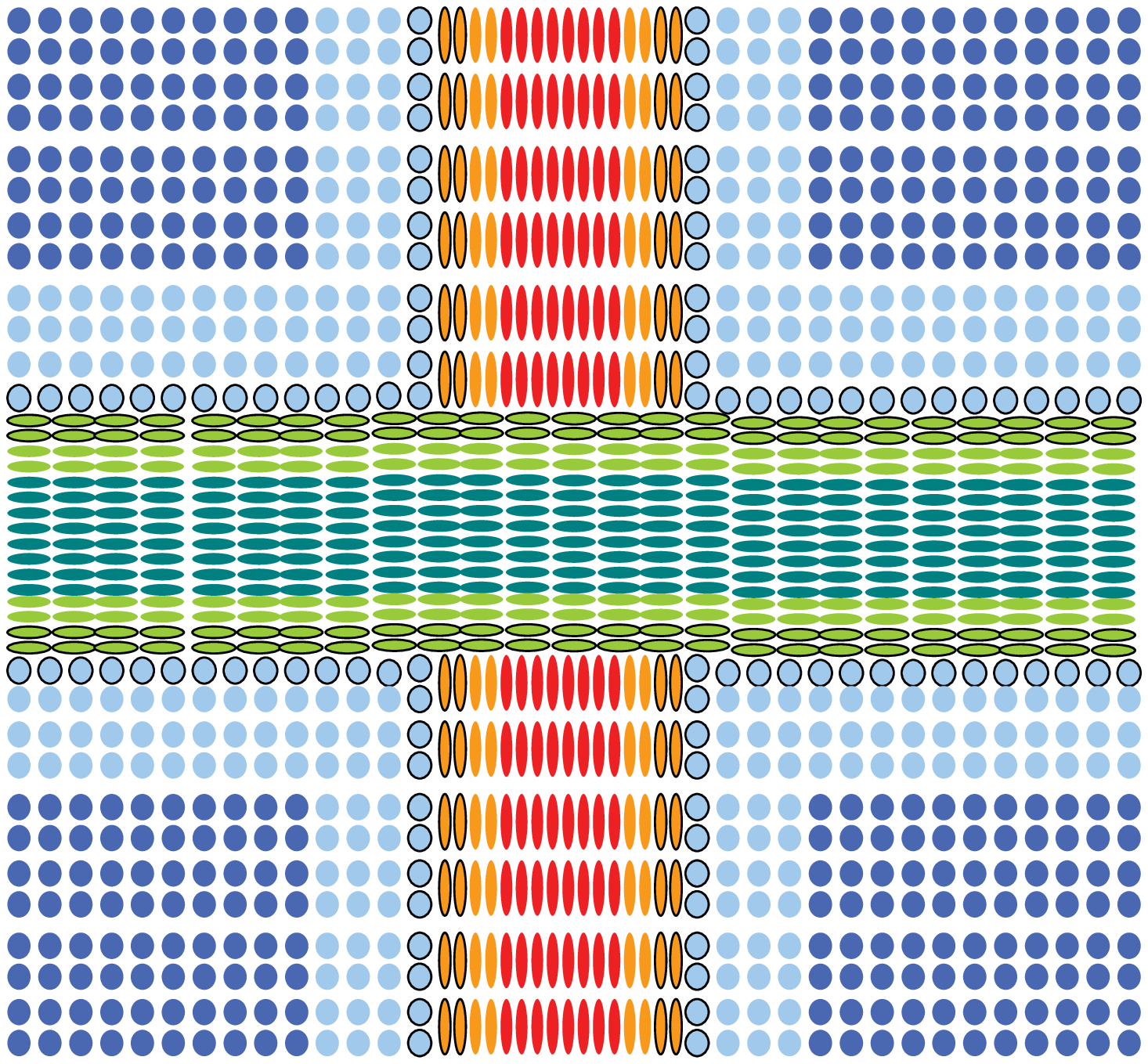}
\caption{Schematic plot of different regions for the simulation
design.} \label{fig:wide_band_regions}
\end{center}
\end{figure}

For a clearer comparison of different smoothing methods, the
tensor fields are divided into different regions and the median
errors within each region are reported for different smoothers and
different bandwidth choices. Six regions are considered and they are
illustrated in the schematic plot Figure
\ref{fig:wide_band_regions}. Specifically, ``whole set'' refers to
the whole tensor field; ``bands crossing'' refers to crossing of two
bands and the boundary of a band and the background (tensors circles
by black line in Figure \ref{fig:wide_band_regions}); ``background interior'' refers to regions within the
background which  are at least four tensors away from any band (dark
blue tensors); ``bands interior'' refers to regions within a band
which are at least four tensors away from the background (red and green
tensors); ``background boundary'' refers to regions within the
background which are within four tensors away of a band (light blue
tensors); and ``bands boundary'' refers to regions on bands which are
within four tensors away from the background (orange and light green tensors). As
mentioned earlier, the background consists of isotropic tensors and
the bands consist of anisotropic tensors. Moreover, both
``background interior'' and ``bands interior'' are relatively
homogeneous regions (i.e., the neighboring tensors are alike),
whereas ``background boundary'', ``bands boundary'' are relatively
heterogeneous regions (i.e., neighboring tensors could be very
different),  and ``bands crossing'' are highly heterogeneous
regions.

\subsubsection*{Results under Rician noise model}\label{subsec:Rician_gradients}

Three smoothers  corresponding to three geometries on the tensor space are
considered, namely (a) Euclidean smoothing; (b) Log-Euclidean
smoothing; and (c) affine-invariant smoothing. Tensors from both
linear regression of observed DWI's (referred to as linear input)
and nonlinear regression of observed DWI's (referred to as nonlinear
input) are used as inputs for tensor smoothing.
We consider both the isotropic smoothing scheme  and the anisotropic
smoothing scheme. In the
latter case, the smoothing is done in two stages. In the first
stage, isotropic smoothing is performed to get a preliminary estimate
$\widehat{D}_{iso}(s)$ at each voxel $s$ and in the second stage,
anisotropic weights are derived by setting
$\widehat{D}=\widehat{D}_{iso}(s)$ in (\ref{eq:weight_aniso_trace})
and then smoothing is performed using the isotropically smoothed
tensors $\widehat{D}_{iso}(s)$'s and these weights. Three
different bandwidth choices are considered for isotropic smoothing:
$h=0.005$, 0.01 and 0.025. For anisotropic smoothing, the bandwidth
pairs being considered are $(h_{iso}, h_{aniso})= (0.005,0.01)$,
$(0.01,0.01)$ and $(0.01,0.025)$, where $h_{iso}$ and $h_{aniso}$,
denote the bandwidths used in the first stage and the second stage,
respectively. We use a Gaussian kernel and truncate the weights that
fall below a small threshold ($10^{-6}$) to speed up computation.
We then re-scale the weights such that their summation equals to one.
The set of neighboring voxels that receive positive weights
is defined as the  neighborhood. The neighborhood does not vary
across tensors for isotropic smoothing at each given bandwidth,
but it is varying for anisotropic smoothing. In  Table 2,
the size of each neighborhood, the number of voxels which account
for (cumulatively) $99\%$ of the total weights (in parentheses),
as well as summary statistics of the weights distribution such as
minimum, median, maximum and entropy (defined as $-\sum_i
\omega_i \log \omega_i$) for isotropic smoothing are reported.
Note that, as bandwidth increases, the entropy of the weights also increases.
Also note that, $h=0.005$ results in essentially no smoothing
(since $>99\%$ of the weights are on the voxel under smoothing itself).

\begin{table}[h]\label{table:kernel_weights}
\begin{center}
\caption{ Neighborhood size and summary statistics of the weights distribution for isotropic smoothing.}
\begin{tabular}{|c|c|cccc|}
\hline
$h$ & Size& min($\omega$) & median($\omega$) & max($\omega$)& Entropy($\omega$) \\
\hline 0.005 & 5 (1) & 0.000881 & 0.000881 & 0.996477 & 0.0283 \\
0.01 &  23 (9)& 0.000002 & 0.000487 & 0.551461 & 1.5140 \\
0.025 & 147 (113) &0.000061 & 0.002371 & 0.071480 & 4.0034 \\
\hline
\end{tabular}
\end{center}
\end{table}

We also conduct simulations where the nine gradient
directions are used only once in the DWI data generation.  Table 3 
reports the median and MAD of the estimation errors for linear and
nonlinear regression procedures across different
regions. It shows that (i) nonlinear regression of the observed DWI data gives more accurate
estimate of the tensors compared to the linear regression. The improvement is only slight on the background (i.e., the isotropic regions). However, the improvement is significant on the bands (i.e., the anisotropic regions); (ii)
comparison under the two
simulation schemes (9 gradients each used once and twice,
respectively) shows that the  results for both linear and nonlinear
regression improve when the number of gradient directions increases;
(iii) results across different noise levels show that the performance of both linear and nonlinear regression degrades
when the noise level increases.

\begin{table}[h]\label{table:nonlinear_vs_linear}
\begin{center}
\caption{Comparison between linear regression and nonlinear
regression on DWI data. Reported numbers are the medians of the
error norms (in parenthesis, MAD of error norms) in three regions:
whole set, bands and background.}
\begin{tabular}{c|c|ccc}
\hline \multicolumn{5}{c}{gradient directions each used
once}\\ \hline
Noise level & Method& Whole set & Bands & Background\\
\hline
$\sigma=0.1$ & linear&  0.1049   (0.0460)  &  0.3207 (0.2409)&       0.0757 (0.0185)\\
& nonlinear&   0.0991   (0.0380)  &   0.1823  (0.1177)       & 0.0757 (0.0185)\\
\hline
$\sigma=0.5$ & linear&    0.5462 (0.2523) &     1.6672  ( 1.3299)&               0.3850  (0.0982)\\
& nonlinear&        0.5141  (0.2148)   &  1.0617    (0.7383)      &     0.3829  (0.0970)\\
\hline

$\sigma=1$ & linear&    1.2382 (0.6738)    &  10.9 ( 5.844)       &        0.819  (0.2592)\\
& nonlinear&                   1.1318  (0.5615)    & 2.8713  (2.3365)&           0.8009  (0.2424)\\
\hline\hline

\multicolumn{5}{c}{9 gradient directions each repeated twice}\\
\hline
Noise level & Method& Whole set & Bands & Background\\
\hline
$\sigma=0.1$ & linear& 0.073891  (0.0322) &   0.229592   (0.1734) &  0.053692  (0.0130)\\
& nonlinear&   0.069904  (0.0267)  &  0.129959    (0.0844)&               0.053679  (0.0130)\\
\hline
$\sigma=0.5$ & linear&   0.383068  (0.1770)    &    1.330171   (1.0666) &            0.271789 (0.0681)\\
& nonlinear&        0.359311   (0.1481)    &    0.828572    (0.5926)   &      0.269491  (0.0672)\\
\hline

$\sigma=1$ & linear&  0.825685 (0.4091)     &      2.489265  (2.0479)&                    0.566317 (0.1579)\\
& nonlinear&                   0.758624  (0.3443)       &    1.726173    (1.2410)&                0.548341  (0.1484)\\
\hline
\end{tabular}
\end{center}
\end{table}


The median errors for different smoothers  across different bandwidths within each of the six regions (illustrated by Figure 2) are plotted in Figures 3 to 8.
In these figures, the $x$
axis denotes bandwidths, and the median errors for
different smoothers and the observed tensors are represented by different symbols, with circle
corresponding to the observed tensor (either derived by linear regression or
by nonlinear regression), triangle corresponding to Euclidean
smoothing,  $+$ corresponding to log-Euclidean smoothing, and
$\times$ corresponding to affine-invariant smoothing.   The main
findings are summarized below.
\begin{enumerate}

\item
Tensor smoothing generally improves upon the corresponding
regression results (either linear or nonlinear) except for the ``bands crossing" regions which are highly heterogenous.

\item
At small noise levels ($\sigma = 0.1$), the three smoothers --
Euclidean, log-Euclidean and affine-invariant -- perform comparably,
with geometric methods working slightly better.

\item
At moderate to large noise levels ($\sigma = 0.5$ or 1), the
Euclidean smoother works better or even considerably better (e.g.
when $\sigma = 1$) than the two geometric smoothers.

\item
The log-Euclidean smoother appears to be more sensitive to noise
than the affine-invariant smoother indicated by a better performance
of the latter under large noise levels ($\sigma=1$). Overall, the
qualitative performances of these two methods are comparable.

\item
Euclidean smoothing results in prominent swelling effects near
the crossing of two bands.


\item
At small noise levels ($\sigma=0.1$), geometric smoothing methods
perform better than Euclidean smoothing in the regions with high
degrees of heterogeneity, i.e., the ``bands crossing'' regions.

\item
At comparable bandwidths,  anisotropic smoothing often improves over
isotropic smoothing. For example, anisotropic with
$(0.01,0.01)$ performs better than isotropic with $h=0.01$ except for  on
``bands crossing''; and  anisotropic with $(0.01,0.025)$ performs
better than isotropic with $h=0.025$ at small noise levels (except for
on ``background boundary'').

\item
In terms of bandwidths choices,  in more homogeneous regions, relatively larger bandwidths are
preferred, whereas in more heterogeneous regions, smaller bandwidths are preferred.
Also, larger bandwidths are preferred under higher noise levels.
More specifically,  on ``bands crossing'' (highly
heterogeneous),  for small noise levels ($\sigma=0.1$), isotropic
smoothing with $h=0.005$ is the best which indeed corresponds to nearly no
smoothing, whereas, for larger noise levels
($\sigma=0.5$, 1), isotropic smoothing with $h=0.01$ is the best. On
``background interior'' (highly homogeneous and isotropic),
anisotropic smoothing with $(0.01, 0.025)$ and isotropic smoothing with $h=0.025$ work
comparably and better than other bandwidth choices. On ``bands interior''
(relatively homogenous and highly anisotropic), anisotropic smoothing with
$(0.01, 0.01)$ or $(0.01, 0.025)$ is the best under small noise
levels, and isotropic smoothing with $h=0.025$ is the best under larger noise
levels. On ``bands boundary'' (relatively heterogeneous),
anisotropic smoothing with $(0.005,0.01)$ is the best under small noise
levels, whereas isotropic smoothing with $h=0.025$ is the best under larger
noise levels. On ``background boundary'' (relatively heterogeneous), under small noise levels,
isotropic smoothing with $h=0.005$ or 0.01 and anisotropic smoothing with $(0.005,0.01)$ or
$(0.01,0.01)$ all work comparably and are better than other
bandwidths choices, whereas under larger noise
levels, anisotropic smoothing with  $(0.01,0.01)$ is the best.

\end{enumerate}

\subsubsection*{Results under spectral noise}

The three smoothers corresponding to { Euclidean, Log-Euclidean, and affine-invariant
metrics are applied to the noise corrupted tensors. Four different bandwidth
choices for isotropic smoothing: $h = 0.005$, $0.01$, $0.025$ and
$0.035$ are considered. For anisotropic smoothing, the bandwidth
pairs being considered are $(h_{iso}, h_{aniso}) = (0.005, 0.01)$
and $(0.01, 0.025)$.

The median errors are plotted in Figures 9 to 14.
The main findings are summarized below.
\begin{enumerate}
\item
At all noise levels, the geometric smoothers outperform the
Euclidean smoother  across all regions except for  ``background
interior'' (highly homogeneous and isotropic)
where Euclidean smoother works slightly better under
larger noise levels $(\nu=20)$.

\item
Log-Euclidean smoother performs slightly better than the
affine-invariant smoother, even though their qualitative
performances are comparable.

\item
Euclidean smoother results in prominent swelling effects near
the crossing of two bands.

\item
Geometric smoothing are more advantageous in the regions that are more
heterogeneous. This can be seen by comparing the results on ``background boundary'' with
``background interior'', and  ``bands boundary'' with ``bands
interior''.

\item
Anisotropic smoothing performs better than isotropic smoothing (at a
comparable bandwidth and especially under smaller noise levels) in the
regions with higher degrees of anisotropy. This is prominent in the
regions  ``bands boundary'' and ``bands interior'' (which together
form the bands).

\item
In terms of bandwidths choices,  in more homogeneous regions,
relatively larger bandwidths are preferred, whereas in
more heterogeneous regions, smaller bandwidths are preferred.
Also, larger bandwidths are preferred under higher noise levels. More specifically,
on ``bands crossing'' (highly heterogeneous), anisotropic
with $(0.005, 0.01)$ and isotropic smoothing with
$h=0.01$ work best; on ``background interior'' (highly homogeneous and
isotropic), anisotropic smoothing with (0.01,0.025), and isotropic
smoothing with $h=0.025, 0.035$ work comparably and better than
other bandwidths choices. On both ``bands interior'' (relatively homogenous and highly anisotropic),
and ``bands boundary'' (relatively heterogeneous and highly
anisotropic), anisotropic smoothing with (0.01, 0.025) works best. On
``background boundary'' (relatively heterogeneous), under smaller
noise levels $(\nu=50)$, anisotropic smoothing with $(0.005,0.01)$ and
$(0.01,0.025)$, and isotropic smoothing with $h=0.01$ work comparably and better
than other bandwidths, whereas, under larger noise levels $(\nu=20)$,
anisotropic smoothing with $(0.01, 0.025)$ works best.

\item
As $\eta$ increases (corresponding to more variable eigenvectors), for any
given $\nu$, the performances of all smoothers across all regions
degrade, except for the ``background interior'' regions
(highly homogeneous and isotropic).
\end{enumerate}

\section{Perturbation analysis of smoothing
methods}\label{sec:perturbation_smoothing}

In this section, we conduct a perturbation analysis to compare the
means of a set of tensors with respect to  the Euclidean,
log-Euclidean and Affine invariant metrics. Arsigny \textit{et al.}
(2005) also carry out a theoretical analysis studying and comparing
log-Euclidean and affine-invariant metrics in tensor computation.
 They derive an asymptotic
expression of the difference between the log-Euclidean mean and the
affine-invariant mean when the true tensors lie in a small
neighborhood of the identity matrix. They also explore situations
where these two means are equal.


We carry out a perturbation analysis by adopting a
framework similar to that in Arsigny \textit{et al.} (2005).
However, we focus mainly on the quantitative aspects  of the
differences between these means. In particular, we derive asymptotic
expansions  for the difference between the Euclidean and
log-Euclidean means, and  that between the Euclidean and affine-invariant means.
We derive these results for general tensors, i.e.,
without requiring them to be nearly isotropic. These results also
highlight the distinction between the situations when the noise
structure is additive (meaning that the
expectation of the tensors equals to the underlying noiseless tensor) and when the
noise structure is non-additive.

Suppose that we observe random tensors  $\{S(\omega) \in \mathcal{P}_N: \omega \in \Omega\}$, where $\Omega$ is an
arbitrary index set with a Borel
$\sigma$-algebra. Let $\mathbb{P}$ be a
probability measure on $\Omega$.

 Let $\bar{S}$ denote the {\it expectation} of $S(\omega)$ with respect to $\mathbb{P}$, i.e.,
\begin{equation}\label{eq:S_bar_Euclidean}
\bar{S} := \mathbb{E}(S) = \int S(\omega) d\mathbb{P}(\omega),
\end{equation}
where the integration is defined through (ordinary) matrix addition of the tensors.
Let $B(\omega) := S(\omega) - \bar{S}$. Then $\mathbb{E}(B) =
\mathbf{0}$, where $\mathbf{0}$ denotes the $N \times N$ zero matrix.

Observe that, the expectation of $S(\omega)$ coincides with
the weighted Karcher mean of $S(\omega)$ with respect to the Euclidean metric,
i.e.,
\begin{equation*}
\bar{S}=\mathbb{E}(S) = \arg\min_{K \in {\cal P}_N} \int d^2(K,S(\omega))
dP(\omega),
\end{equation*}
where $d(\cdot,\cdot)$ is a Euclidean distance. Next, let
$\bar{S}_{LE}$ denote the mean of $S(\omega)$ with
respect to the log-Euclidean metric, i.e.,
\begin{equation}\label{eq:log_LE_mean}
\bar{S}_{LE}  :=  \arg\min_{K \in {\cal P}_N} \int
d_{LE}^2(K,S(\omega)) dP(\omega).
\end{equation}
It is easy to show that,
$$
\log \bar{S}_{LE} = \mathbb{E}(\log S)=\int \log S(\omega)
dP(\omega).
$$
Finally, let $\bar{S}_{Aff}$ denote the affine-invariant mean of $S(\omega)$, i.e.,
\begin{equation}\label{eq:affine_S_mean}
\bar{S}_{Aff} := \arg\min_{K \in {\cal P}_N} \int
d_{Aff}^2(K,S(\omega)) dP(\omega),
\end{equation}
In general, there is no closed form expression for
$\bar{S}_{Aff}$, but $\bar{S}_{Aff}$ satisfies the following
barrycentric equation (Arsigny \textit{et al.}, 2005):
\begin{equation}\label{eq:affine_S_mean_barycentric}
\mathbb{E}(\log(\bar{S}_{Aff}^{-1/2} S \bar{S}_{Aff}^{-1/2})) = \mathbf{0}.
\end{equation}

In the following, we adopt a matrix perturbation analysis approach
to study the differences among  means under different metrics. Let
$\lambda_j$ denote the $j$-th largest \textit{distinct} eigenvalue
of $\bar{S}$ and let $P_j$ denote the corresponding
eigen-projection. We make a distinction between the cases when
$\bar{S}$ is a multiple of the identity (i.e., $\bar{S}$ is isotropic) and when otherwise (i.e., $\bar{S}$ is anisotropic). Let
$$
C := \begin{cases}
\max\{\max_j \lambda_j^{-1},\max_{k \neq j}|\lambda_k - \lambda_j|^{-1}\}, & \mbox{if}~\bar{S} \neq \lambda_1 I \\
\lambda_1^{-1} & \mbox{if}~\bar{S} = \lambda_1 I \\
\end{cases}
$$
where  $k \neq j$ corresponding to pairs of distinct eigenvalues of $\bar{S}$ and
$I$ denotes the $N \times N$ identity matrix. We also assume that
the probability measure $\mathbb{P}$ satisfies
\begin{equation}\label{eq:eigen_perturb_cond}
\mathbb{P}(\sup_{\omega \in \Omega}
\parallel B(\omega)
\parallel < C^{-1}t) = 1,
\end{equation}
for some $t>0$, where $\parallel \cdot \parallel$ denotes the operator
norm. Note that, $t$ can be viewed as a scale parameter indicating
the spread of $S(\omega)$'s. A small value of $t$ means
that the tensors are more homogeneous.

In the case that $\bar{S}$ is not a multiple of the identity (i.e.,
$\bar{S}$ is anisotropic), define $H_j$ to be the matrix
\begin{equation}\label{eq:H_j_def}
H_j := \sum_{k \neq j} \frac{1}{\lambda_k - \lambda_j} P_k.
\end{equation}
Note that $H_j P_j = P_j H_j = \mathbf{0}$ for all $j$. For
simplicity of exposition, in the following,  we consider only two cases: (a) when the
eigenvalues of $\bar{S}$ are all distinct; and (b) when $\bar{S}$ is
isotropic, i.e., all its eigenvalues are equal to $\lambda_1$.

\subsection*{Comparison of Euclidean and log-Euclidean
means}\label{subsec:E_vs_LE}

\begin{proposition}\label{prop:E_vs_LE}
Suppose that the tensors $\{S(\omega) : \omega \in \Omega\}$ and the
probability distribution $\mathbb{P}$ satisfy
(\ref{eq:eigen_perturb_cond}).
\begin{itemize}
\item[(a)]
If the eigenvalues of $\bar{S} = \mathbb{E}(S)$ are all distinct,
then
\begin{eqnarray}\label{eq:log_S_mean_expansion}
&& \log \bar{S}_{LE} - \log \bar{S} \nonumber\\
&=& - \sum_{j=1}^N \frac{1}{\lambda_j} \mathbb{E}[\tr(P_j B H_j B)]
P_j - \frac{1}{2}\sum_{j=1}^N \frac{1}{\lambda_j^2} \mathbb{E}
[\tr(P_j B)]^2 P_j \nonumber\\
&& + \sum_{j=1}^N \log \lambda_j \mathbb{E}\left[ P_j B H_j B H_j +
H_j B P_j B H_j + H_j B H_j B P_j \right. \nonumber\\
&& \left. - P_j B P_j B H_j^2 - P_j B H_j^2
B P_j - H_j^2 B P_j B P_j\right] \nonumber\\
&& - \sum_{j=1}^N \frac{1}{\lambda_j} \mathbb{E}[\tr(P_j B)  (P_j B
H_j + H_j B P_j)] + O(t^3).
\end{eqnarray}

\item[(b)] If $\bar{S} = \lambda_1 I$, then
\begin{equation}\label{eq:log_S_mean_expansion_isotropic}
\log \bar{S}_{LE} - \log\bar{S} = -
\frac{1}{2\lambda_1^2}\mathbb{E}(B^2) + O(t^3).
\end{equation}
\end{itemize}
\end{proposition}

\begin{remark}\label{rem:logdet_E_vs_LE}
When the eigenvalues of $\bar{S}$ are all distinct, taking trace on
both sides of (\ref{eq:log_S_mean_expansion}), and using the fact
that $P_j H_j = \mathbf{0}$ and $P_j^2 = P_j$, we get
\begin{eqnarray}\label{eq:trace_log_S_mean_expansion}
\tr(\log \bar{S}_{LE}) - \tr(\log \bar{S}) &=& - \sum_{j=1}^N
\frac{1}{\lambda_j} \mathbb{E}[\tr(P_j B H_j B )] -
\frac{1}{2}\sum_{j=1}^N \frac{1}{\lambda_j^2} \mathbb{E} [\tr(P_j
B)]^2 +  O(t^3).
\end{eqnarray}
Since
$$
\tr(\log \bar{S}_{LE}) - \tr(\log \bar{S}) = \log\det(\bar{S}_{LE})
- \log\det(\bar{S}),
$$
where $\det$ denotes the determinant,
(\ref{eq:trace_log_S_mean_expansion}) actually gives an expansion of
the determinant of $\bar{S}_{LE}$ around the determinant of
$\bar{S}$.
\end{remark}

\subsection*{Comparison of Euclidean and affine-invariant
means}\label{subsec:E_vs_Affine}

\begin{proposition}\label{prop:E_vs_Affine}
Suppose that the tensors $\{S(\omega) : \omega \in \Omega\}$ and the
probability distribution $\mathbb{P}$ satisfy
(\ref{eq:eigen_perturb_cond}).
\begin{itemize}
\item[(a)]
If the eigenvalues of $\bar{S}$ are all distinct, then
\begin{eqnarray}\label{eq:log_S_affine_mean_expansion}
&& \log \bar{S}_{Aff} -\log \bar{S} \nonumber\\
&=&  \sum_{j=1}^N \frac{1}{\lambda_j} \tr(P_j(\bar{S}_{Aff} -
\bar{S})) P_j - \sum_{j=1}^N \log\lambda_j \left(P_j(\bar{S}_{Aff} -
\bar{S}) H_j + H_j
(\bar{S}_{Aff} - \bar{S}) P_j\right) + O(t^4) \nonumber\\
&=& - \frac{1}{2} \sum_{j=1}^N \frac{1}{\lambda_j}
\mathbb{E}[\tr(P_j B \bar{S}^{-1} B)] P_j \nonumber\\
&& ~~ + \frac{1}{2} \sum_{j=1}^N \log\lambda_j \mathbb{E}[P_j B
\bar{S}^{-1} B H_j + H_j B \bar{S}^{-1} B P_j] + O(t^3).
\end{eqnarray}

\item[(b)]
If $\bar{S} = \lambda_1 I$, then
\begin{equation}\label{eq:log_S_affine_mean_expansion_isotropic}
\log \bar{S}_{Aff} - \log\bar{S} = -
\frac{1}{2\lambda_1^2}\mathbb{E}(B^2) + O(t^3).
\end{equation}
\end{itemize}
\end{proposition}

Note that, Propositions 1 and 2 can be used together to give an asymptotic expansion
for the difference between the log-Euclidean and affine-invariant
means.

\begin{remark}\label{rem:logdet_E_vs_Affine}
When the eigenvalues of $\bar{S}$ are all distinct, taking trace on
both sides of (\ref{eq:log_S_affine_mean_expansion}), and using $P_j
H_j = \mathbf{0}$, we get
\begin{equation}\label{eq:trace_log_S_mean_affine_expansion}
\log \det(\bar{S}_{Aff}) - \log \det(\bar{S}) = \tr(\log
\bar{S}_{Aff}) - \tr(\log \bar{S}) = - \frac{1}{2} \sum_{j=1}^N
\frac{1}{\lambda_j} \mathbb{E}[\tr(P_j B \bar{S}^{-1} B)] + O(t^3).
\end{equation}
\end{remark}

\begin{remark}\label{rem:difference}
Expressions (\ref{eq:log_S_mean_expansion_isotropic}) and
(\ref{eq:log_S_affine_mean_expansion_isotropic}) show that, when
$\bar{S} = \lambda_1 I$ (i.e., isotropic), $\bar{S}_{LE}$ and
$\bar{S}_{Aff}$ are the same up to the second order (in terms of the
dispersion parameter $t$). On the other hand, from
(\ref{eq:log_S_mean_expansion}) and
(\ref{eq:log_S_affine_mean_expansion}) we can conclude that when
$\bar{S}$ is sufficiently far from being isotropic, there is a second
order difference between $\bar{S}_{LE}$ and $\bar{S}_{Aff}$.
Moreover, in these expansions, the terms involving $\log \lambda_j$
tend to dominate when at least one of the eigenvalues of $\bar{S}$
is close to zero. Thus, we expect to see a large difference between
the log-Euclidean and affine-invariant means when $\bar{S}$ is
anisotropic and has near zero eigenvalues. This is also confirmed by
simulations.
\end{remark}

\begin{remark}\label{rem:additive_noise}
When the noise is {\it additive}, i.e. $\bar{S} = \mathbb{E}(S) = S^*$
where $S^*$ denotes the ``true'' or target tensor, our analysis
shows that both $\bar{S}_{Aff}$ and $\bar{S}_{LE}$ are biased
``estimators'' for $S^*$ and the bias can be quantified by
Propositions \ref{prop:E_vs_LE} and \ref{prop:E_vs_Affine}.
\end{remark}

\begin{remark}\label{rem:multiplicative_noise}
Suppose that the noise is multiplicative in the following sense. Let
$S^* = G \Delta G^T$ be the true tensor where the columns of the
$N\times N$ orthogonal matrix $G$ denote the eigenvectors
corresponding to the eigenvalues $\delta_1 \geq \cdots \geq \delta_N
>0$, which are the diagonal elements of the diagonal matrix
$\Delta$. Suppose that the observed tensors are of the form
$$
S(\omega) = G ~\mbox{diag}(\delta_1 e^{Z_1(\omega)},\cdots,\delta_N
e^{Z_N(\omega)}) G^T
$$
where $Z_1,\ldots,Z_N$ are independent random variables with $Z_j$
having uniform distribution on the interval $[-c_jt,c_j t]$ for
constants $c_1,\ldots,c_N > 0$. Then, the tensors $S(\omega)$
satisfy (\ref{eq:L_sup_order}) (see the proof of proposition 2). Moreover, the tensors commute with
each other, which implies that $\bar{S}_{LE} = \bar{S}_{Aff}$. Now,
using the fact that $\mathbb{E}(Z_j) = 0$ ($j=1,\ldots,N$), we have
\begin{eqnarray*}
\bar{S}_{LE} &=& \exp(G ~\mbox{diag}(\mathbb{E}(\log \delta_1 +
Z_1),\cdots,\mathbb{E}(\log \delta_N + Z_N)) G^T)
\\
&=& \exp(G \log\Delta G^T) = G \Delta G^T = S^*.
\end{eqnarray*}
However, $\mathbb{E}(e^{Z_j}) = (e^{c_jt} - e^{-c_jt})/(2c_j t) \neq
1$, which implies that $\bar{S} \neq S^*$. In this case, the
Euclidean mean will be a biased estimator for the true tensor $S^*$,
and the bias is of the order $O(t^2)$.
\end{remark}

\section{Regression analysis of DWI data under Rician noise model}\label{sec:reg_DWI}

In this section, we study the {\it Rician noise model} for diffusion weighted imaging data (Hanh \textit{et
al.}, 2006). Our goal is to
compare two regression procedures for deriving diffusion tensors from raw
diffusion weighted images under  this noise model.

In DT-MRI studies, the raw data obtained by MRI scanning are complex
numbers representing the Fourier transformation of a magnetization
distribution of a tissue at a certain point in time (cf. Mori,
2007). If the electronic noise in the real and imaginary parts of
the raw data are assumed to be independent Gaussian  random
variables (Henkelman 1985; Gudbjartsson and Patz 1995; Macorski
1996), then the corresponding signal intensities follow a Rician
distribution. Statistical properties of this noise model are studied
by Zhu \textit{et al.} (2009), where diagnostic tools to assess the
quality of MR images are developed. Zhu \textit{et al.} (2007)
propose a semiparametric model for noise in diffusion-weighted
images, and develop a weighted least squares estimate of the
tensors. They study the effects of noise on the estimated diffusion
tensors, as well as on their eigen-structures and morphological
classification. The Rician noise model is also analyzed by Polzehl
and Tabelow (2008) where a weighted version of the maximum
likelihood estimator of the tensors is proposed.

In this section, we consider the Rician noise model and focus on:
(i) obtaining asymptotic
expansions of the linear and nonlinear least squares estimators in
high signal-to-noise ratio (SNR) regimes -- the results show that the linear regression
estimate is less efficient; and (ii) showing that the
nonlinear least squares estimator is asymptotically as efficient as
the maximum likelihood estimator in high SNR regimes;
(iii) quantifying the bias in the linear least squares estimator at low SNR.
Here we assume that the gradient
directions in the MR data acquisition step is fixed and the noise
level varies. In contrast, in Zhu \textit{et al.} (2007, 2009), the
number of gradient directions increases to infinity while the noise
level is kept fixed.

\subsection{Rician noise model and regression estimates}\label{subsec:setup_regression}

We first describe the Rician noise model for diffusion-weighted MR
images at a given voxel.
Let $\mathcal{B}$ denote a set of gradient directions ($3 \times 1$
vectors of unit norm). Under the  model for diffusion
tensors (cf. Mori, 2007), the noiseless diffusion weighted signal at
direction $b \in \mathcal{B}$ is given by
\begin{equation}\label{eq:Rician_signal}
\overline{S}_b = S_0 \exp(-b^T D b),
\end{equation}
where $S_0$ is the baseline intensity, and $D$ is a $3 \times 3$
positive definite matrix (the diffusion tensor). The Rician noise
model says that, the observed diffusion weighted signal $S_b$
is a random variable obtained by:
\begin{equation}\label{eq:S_b_def}
S_b = \left(|\overline{S}_b u_{b1} + \sigma
\varepsilon_{b1}|^2 + |\overline{S}_b u_{b2} + \sigma
\varepsilon_{b2}|^2\right)^{1/2} =
\parallel \overline{S}_b u_b + \sigma \varepsilon_b\parallel,
\end{equation}
where $u_b := (u_{b1},u_{b2})^T$ is a unit vector in $\mathbb{R}^2$,
and the random vectors $\varepsilon_b :=
(\varepsilon_{b1},\varepsilon_{b2})^T$ are distributed as $N(0,I_2)$
and are assumed to be independent for different $b$. The parameter
$\sigma > 0$ controls the noise level.


We consider two regression based methods for estimating $D$ based on
the observed DWI signal intensities $\{S_b: b \in
\mathcal{B}\}$. For simplicity, throughout this section, $S_0$ is
assumed to be known and fixed. The first method is to use the log
transformed DWI's to estimate $D$ by a linear regression. The resulting
estimator is referred to as the \textit{linear regression estimator}
and denoted by $\widehat D_{LS}$. The second method uses nonlinear
regression based on the original DWI's to estimate $D$. The
resulting estimator is referred to as the \textit{nonlinear
regression estimator} and denoted by $\widehat D_{NL}$.

In order to define these estimators, we introduce some notations.
Specifically, the tensor $D$ is treated as a $6\times 1$ vector with
elements $D_{11},D_{22},D_{33},D_{12},D_{13},D_{23}$. Under this
notation, the quadratic form $b^T D b$ (where $D$ is a matrix) can
be rewritten as $x_b^T D$ (where $D$ is a vector) with $x_b =
(b_1^2,b_2^2,b_3^2,2b_1b_2,2b_1b_3,2b_2b_3)^T$. In the following, we
also assume that the matrix $\sum_{b \in {\cal B}} x_b x_b^T$ is
well-conditioned, which is guaranteed by an appropriate choice of
the gradient directions. Then the linear regression and non-linear
regression estimators are defined as
\begin{equation}\label{eq:log_LS}
\widehat D_{LS} := \arg\min_D \sum_{b\in {\cal B}}(\log S_b -
\log S_0 + x_b^T D)^2,
\end{equation}
and
\begin{equation}\label{eq:log_NL}
\widehat D_{NL} := \arg\min_D \sum_{b\in {\cal B}}(S_b - S_0
\exp(-x_b^T D))^2.
\end{equation}
Note that, these estimators are used in the simulation studies under the Rician noise model (Section
\ref{sec:simulation}) to derive the observed tensor at each
voxel,  which is then used as input for the smoothing methods.

In the following subsections, we analyze the behavior of the above
estimators by varying the noise level, i.e., $\sigma$, which
corresponds to varying signal-to-noise ratio (SNR). Consistency of
the estimators is established under the $\sigma \to 0$ setting,
i.e., the  ``large SNR'' scenario.  In addition, we also compare the
nonlinear regression estimator with the maximum likelihood
estimator which utilizes the Rician noise model explicitly.
Throughout, we shall use $D^0$ to denote the true diffusion tensor.

\subsection{Asymptotic analysis of $\widehat D_{LS}$ and $\widehat D_{NL}$ under large SNR}\label{subsec:asymp_regression}

In this subsection, we carry out asymptotic expansions of the
estimates $\widehat D_{LS}$ and $\widehat D_{NL}$ assuming that the
noise level $\sigma$ goes to zero, while $\overline{S}_b$'s are
treated as fixed.  Effectively, this means that, $\sigma
\ll \min_{b \in {\cal B}} \overline{S}_b$, which is equivalent to
assuming that $S_0 \gg \sigma$ and that $\max_{b \in {\cal B}} x_b^T
D^0$ is ``not too large''. Clearly, the last condition holds if the
largest eigenvalue of the tensor $D^0$ is not very large (note that,
the gradient directions are normalized to have unit norm).

\subsubsection*{Asymptotic expansion of $\widehat D_{LS}$}

\begin{proposition}\label{prop:D_hat_LS_asymp}
As $\sigma \to 0$, $\widehat D_{LS} = D^0 + \sigma D_{1,LS} +
\sigma^2 D_{2,LS} + O_P(\sigma^3)$ where the random vectors
$D_{1,LS}$ and $D_{2,LS}$ are given by
\begin{equation}\label{eq:D_1_LS}
D_{1,LS} =  - (\sum_{b \in {\cal B}} x_b x_b^T)^{-1} (\sum_{b \in
{\cal B}} \frac{1}{\overline{S}_b} (u_b^T\varepsilon_b) x_b),
\end{equation}
and
\begin{equation}\label{eq:D_2_LS}
D_{2,LS} = -\frac{1}{2}(\sum_{b \in {\cal B}} x_b x_b^T)^{-1}
(\sum_{b \in {\cal B}} \frac{1}{\overline{S}_b^2}( (v_b^T
\varepsilon_b)^2 - (u_b^T \varepsilon_b)^2) x_b).
\end{equation}
\end{proposition}

\begin{remark}\label{rem:D_hat_LS}
$D_{1,LS}$ has a normal distribution with mean zero  and variance
\begin{equation}\label{eq:var_D_1_LS}
\mbox{Var}(D_{1,LS}) = (\sum_{b \in {\cal B}} x_b x_b^T)^{-1}
(\sum_{b \in {\cal B}} {\overline{S}_b}^{-2} x_b x_b^T)(\sum_{b \in
{\cal B}} x_b x_b^T)^{-1},
\end{equation}
whereas $D_{2,LS}$ is a weighted sum of differences of independent
$\chi_{(1)}^2$ random variables, and hence it also has mean 0. These
show that the bias in $\widehat D_{LS}$ is of the order
$O(\sigma^3)$.
\end{remark}

\subsubsection*{Asymptotic expansion of $\widehat D_{NL}$}

Unlike $\widehat D_{LS}$, there is no explicit expression for
$\widehat D_{NL}$. Instead, it satisfies the normal equation:
\begin{eqnarray}\label{eq:NL_normal}
\sum_{b \in {\cal B}} (S_b - S_0 \exp(-x_b^T \widehat D_{NL}))
S_0 \exp(-x_b^T \widehat D_{NL}) x_b &=& 0.
\end{eqnarray}
We have the following result.
\begin{proposition}\label{prop:D_hat_NL_asymp}
As $\sigma \to 0$, $\widehat D_{NL} = D^0 + \sigma D_{1,NL} +
\sigma^2 D_{2,NL} + O_P(\sigma^3)$, where
\begin{equation}\label{eq:D_1_NL}
D_{1,NL} =  - (\sum_{b \in {\cal B}} \overline{S}_b^2 x_b
x_b^T)^{-1} (\sum_{b \in {\cal B}}  \overline{S}_b (u_b^T
\varepsilon_b) x_b),
\end{equation}
and
\begin{equation}\label{eq:D_2_NL}
D_{2,NL} = (\sum_{b \in {\cal B}} \overline{S}_b^2 x_b x_b^T)^{-1}
\left[\sum_{b \in {\cal B}}  \overline{S}_b^2 (x_b^T D_{1,NL})^2 x_b
+ \frac{1}{2} \sum_{b \in {\cal B}} (\overline{S}_b x_b^T D_{1,NL} +
u_b^T \varepsilon_b)^2 x_b -
\frac{1}{2} \sum_{b \in {\cal B}} \parallel \varepsilon_b \parallel^2 x_b\right]. \\
\end{equation}
\end{proposition}

\begin{remark}\label{rem:D_hat_NL}
$D_{1,NL}$ has a normal distribution with mean zero and variance
\begin{equation}\label{eq:var_D_1_NL}
\mbox{Var}(D_{1,NL}) = (\sum_{b \in {\cal B}} \overline{S}_b^2 x_b
x_b^T)^{-1},
\end{equation}
and $D_{2,NL}$ is a weighted sum of (dependent) $\chi^2_{(1)}$ random
variables, with mean
\begin{equation}\label{eq:mean_D_2_NL}
\mathbb{E}(D_{2,NL}) = - \frac{1}{2} (\sum_{b \in {\cal B}} \overline{S}_b^2
x_b x_b^T)^{-1} \left[ \sum_{b\in {\cal B}} \left(1-\overline{S}_b^2
x_b^T (\sum_{b' \in {\cal B}} \overline{S}_{b'}^2 x_{b'}
x_{b'}^T)^{-1} x_b\right) x_b\right].
\end{equation}
Observe that in (\ref{eq:mean_D_2_NL}), the coefficients of all the
$x_b$'s, are negative  and thus the mean of $D_{2,NL}$ is
non-vanishing. Therefore, the bias in $\widehat D_{NL}$ is of order
$O(\sigma^2)$.
\end{remark}

\begin{remark}\label{rem:Rician_additive}
Propositions \ref{prop:D_hat_LS_asymp} and
\ref{prop:D_hat_NL_asymp} show that, under the Rician noise model, at relatively high SNR, both
linear and nonlinear regression estimators of the diffusion tensor
are nearly unbiased. This means that under such a scenario, when the regression estimators are used as input for the smoothing methods,  the corresponding noise structure is nearly additive.
\end{remark}


\subsubsection*{Comparison of $\widehat D_{LS}$ and $\widehat D_{NL}$}

It is clear from the above expansions that both $\widehat D_{LS}$
and $\widehat D_{NL}$ are consistent estimators when $\sigma$ goes
to zero, where $\widehat D_{LS}$ has a bias of order $O(\sigma^3)$
and that in $\widehat D_{NL}$ is of order $O(\sigma^2)$. Up to the
first order, both estimators have Gaussian fluctuations, and up to
the second order, they become weighted sums of Gaussian and
chi-squared random variables. The asymptotic analysis also points to
an important advantage of $\widehat D_{NL}$ over
$\widehat D_{LS}$: the former has a smaller asymptotic variance than
the latter. That is,
\begin{equation}\label{eq:asymp_var_D_NL_LS_compare}
(\sum_{b \in {\cal B}} \overline{S}_b^2 x_b x_b^T)^{-1} \preceq
(\sum_{b \in {\cal B}} x_b x_b^T)^{-1} (\sum_{b \in {\cal B}}
{\overline{S}_b}^{-2} x_b x_b^T)(\sum_{b \in {\cal B}} x_b
x_b^T)^{-1}
\end{equation}
where $\preceq$ denotes the inequality between positive
semi-definite matrices. This can be proved by using Schur complement
and noticing that the matrix
$$
\begin{bmatrix}
\overline{S}_b^2 x_b x_b^T & x_b x_b^T \\
x_b x_b^T & \overline{S}_b^{-2} x_b x_b^T \\
\end{bmatrix}
$$
is positive semi-definite for each $b \in {\cal B}$. Indeed, the
difference between the two asymptotic covariance matrices can be
quite substantial if some $\overline{S}_b$'s are significantly
smaller than  $(1/|\mathcal{B}|) \sum _{b \in
\mathcal{B}}\overline{S}_b$. This situation can arise when the true
diffusion tensor $D^0$ has a strong directional component (i.e.,
being a highly anisotropic tensor). This is because, under such a
situation,  only a few gradient directions are likely to be aligned
to the leading eigen-direction, which results in small
$\overline{S}_b$, whereas other gradient directions lead to large
$\overline{S}_b$. Consequently,  $\widehat D_{LS}$ will have a high
degree of variability and the quadratic forms $b^T D b$ associated
with the gradient directions that are partially aligned to the
leading eigen-direction will be poorly estimated. This fact has also
been empirically verified through simulation studies (results not
reported here).

\subsection{Comparison of MLE and $\widehat D_{NL}$ under large SNR}
\label{subsec:Fisher_Rician}

The probability density function of the Rician noise distribution
with signal parameter $\zeta$ (representing the amplitude of the
signal) and noise parameter $\sigma$ is given by (cf. Polzehl and
Tabelow, 2008)
\begin{equation}\label{eq:Rician_pdf}
p_{\zeta,\sigma}(x) = \frac{x}{\sigma^2}
\exp\left(-\frac{x^2+\zeta^2}{2\sigma^2}\right)
I_0\left(\frac{x\zeta}{\sigma^2}\right) \mathbf{1}(x > 0),
\end{equation}
where $I_0$ denotes the zero-th order modified Bessel function of
the first kind. Under the Rician noise model
(\ref{eq:Rician_signal}) and (\ref{eq:S_b_def}), the measurements
$S_b$'s are independent with p.d.f. $p_{\zeta(D^0,b),\sigma}$,
where $D^0$ is the true tensor and
\begin{equation}\label{eq:zeta_D_b_def}
\zeta(D,b) :=  S_0 \exp(-x_b^T D)
\end{equation}
so that $\overline{S}_b = \zeta(D^0,b)$.

We first compute the Fisher information matrix for the Rician noise
model with respect to $D$. In the following, for simplicity,
$\sigma$ is treated as known.
\begin{proposition}\label{prop:Fisher_Rician}
Under model (\ref{eq:Rician_signal}) and (\ref{eq:S_b_def}), for a given $\sigma>0$, the Fisher information matrix with respect
to $D$ at $D^0$ based on the data $\{S_b : b \in {\cal B}\}$,
is
\begin{eqnarray}\label{eq:Fisher_D_0}
{\cal I}(D^0;\sigma) &:=& \sum_{b \in {\cal B}}
\mathfrak{I}(\zeta(D^0,b);\sigma) (\zeta(D^0,b))^2 x_b x_b^T \nonumber\\
&=& \sum_{b \in {\cal B}}\mathfrak{I}(\overline{S}_b;\sigma)
\overline{S}_b^2 x_b x_b^T
\end{eqnarray}
Here $\mathfrak{I}(\zeta;\sigma)$ is the Fisher information for the
parameter  $\zeta$ of the Rician distribution (\ref{eq:Rician_pdf}),
and it can be expressed as
\begin{equation}\label{eq:Fisher_Rician}
\mathfrak{I}(\zeta;\sigma) = \frac{1}{\sigma^2}\left[
e^{-\frac{\zeta^2}{2\sigma^2}} V\left(\frac{\zeta}{\sigma}\right)-
\frac{\zeta^2}{\sigma^2}\right],
\end{equation}
where, for $w > 0$,
\begin{equation}\label{eq:V_w_expr}
V(w) := \int_0^\infty u^3 \frac{(I_1(uw))^2}{I_0(uw)} e^{-u^2/2} du
= \frac{1}{w^4} \int_0^\infty v^3 \frac{(I_1(v))^2}{I_0(v)}
e^{-v^2/(2w^2)} dv.
\end{equation}
\end{proposition}
\begin{remark}\label{rem:V_w_expr}
There is no closed form expression for $V(w)$. However, since both
$I_1$ and $I_0$ are monotone increasing $C^\infty$ functions with
subexponential growth on $[0,\infty)$, this function can be
evaluated for any given $w$ by numerical integration using an
appropriate quadrature method.
\end{remark}


Let $\widehat D_{ML}$ denote the maximum likelihood estimator of
$D$. Then $\widehat D_{ML}$ satisfies the likelihood equation
\begin{equation}\label{eq:MLE_D}
\sum_{b \in {\cal B}} \nabla \log
p_{\zeta,\sigma}(S_b)\left|_{\zeta = \zeta(\widehat
D_{ML},b)}\right. = 0,
\end{equation}
where $\nabla \log p_{\zeta(D,b),\sigma}(x)$ denotes the gradient of
$\log p_{\zeta(D,b),\sigma}(x)$ with respect to the parameter $D$.
Then, similarly as in the case of $\widehat D_{NL}$,  it can be
shown that, as $\sigma \to 0$, $\widehat D_{ML} = D^0 + \sigma
D_{1,ML} + o_P(\sigma)$, where
\begin{equation}\label{eq:D_ML_first_order}
\mathbb{E}(D_{1,ML}) = 0, \qquad\mbox{and}\qquad
\mbox{Var}(D_{1,ML}) = (\sum_{b \in {\cal B}} \overline{S}_b^2 x_b
x_b^T)^{-1}.
\end{equation}
This implies that the asymptotic variance of $\widehat D_{ML}$ is
$\sigma^2(\sum_{b \in {\cal B}} \overline{S}_b^2 x_b x_b^T)^{-1} +
o(\sigma^2)$.

It can be checked that for every fixed $\zeta$, the quantity
$\sigma^2 \mathfrak{I}(\zeta;\sigma)$ is less than 1 for all $\sigma
> 0$ and converges to 1 as  $\sigma \to 0$. From this, it follows
that
\begin{equation}\label{eq:Fisher_D_0_sigma_limit}
\sigma^2 {\cal I}(D^0;\sigma)  \to \sum_{b \in {\cal B}}
\overline{S}_b^2 x_b x_b^T ~~~\mbox{as}~~~ \sigma \to 0.
\end{equation}
Since $\sigma = o(1)$, by Proposition \ref{prop:D_hat_NL_asymp}, we
have $\widehat D_{NL} = D^0 + \sigma D_{1,NL} + O_P(\sigma^2)$.
Using (\ref{eq:var_D_1_NL}) and (\ref{eq:D_ML_first_order}) it can
be concluded that $\widehat D_{NL}$ has the same asymptotic variance
as the MLE when $\sigma \to 0$. In other words, under the large SNR
regime, the nonlinear regression estimator $\widehat D_{NL}$ is
asymptotically as efficient as the MLE $\widehat D_{ML}$. Moreover,
(\ref{eq:Fisher_D_0_sigma_limit}) shows that when scaled by
$\sigma^2$, the (common) asymptotic covariance matrix of $\widehat
D_{NL}$ and MLE is the  limit of the inverse of the Fisher information
matrix of $D$.

\subsection{Bias in $\widehat D_{LS}$ under small SNR }\label{sec:bias_LS}

Polzehl and Tabelow (2008) discuss in detail the issue of bias in
estimating tensors from the raw DWI data. They show that the raw DWI
signal $S_b$ is a biased estimator of $\overline{S}_b$ and the
bias becomes larger at smaller SNR. Specifically, they derive an
expression for the bias which is quoted below.

Let $B(x_b^T D^0,\sigma) := E_{D^0,\sigma}(S_b) - S_0e^{-
x_b^T D^0}$ denote the bias in $S_b$ as an estimate of
$\overline{S}_b$ when $(D^0,\sigma)$ is the true parameter.  Then
\begin{equation}
B(x_b^T D^0,\sigma) = \sqrt{\frac{\pi}{2}} \sigma
L_{1/2}\left(-\frac{S_0^2e^{-2x_b^T D^0}}{2\sigma^2}\right) -
S_0e^{-x_b^T D^0}
\end{equation}
where $L_{1/2}(x) = e^{x/2}[(1-x)I_0(-x/2) - x I_1(-x/2)]$, and
$I_\nu$ is the $\nu$-th order modified Bessel function of the first
kind (cf. Abramowitz and Stegun, 1965).


In this subsection, we investigate the effect of the noise on
$\widehat D_{LS}$. For simplicity, we  consider an idealized
framework where the ``informative'' gradient directions can be
separated from the ``noninformative'' ones.


Suppose that there is a direction $b \in {\cal B}$ such that $x_b^T
D^0$ is large in the sense that, $\overline{S}_b = S_0 \exp(-x_b^T
D^0) \ll \sigma$, even though $\sigma \ll S_0$. In that case, the
analysis presented in Section \ref{subsec:asymp_regression} is not
valid since the Taylor expansions adopted there is not appropriate
due to the explosive behavior of the higher order derivatives of the
associated functionals. Instead, we consider the following
approximation:
\begin{eqnarray}\label{eq:log_S_b_expand_small_S_b}
\log S_b - \log S_0 &=& \log\parallel e^{-x_b^T D^0} u_b +
\frac{\sigma}{S_0} \varepsilon_b\parallel \nonumber\\
&=& \frac{1}{2} \log\left(\frac{\sigma^2}{S_0^2} \parallel
\varepsilon_b \parallel^2 + 2\frac{\sigma}{S_0} e^{-x_b^T D^0}
u_b^T \varepsilon_b + e^{-2x_b^T D^0}\right) \nonumber\\
&=&\frac{1}{2}\log\left(\frac{\sigma^2}{S_0^2} (\parallel
\varepsilon_b \parallel^2 + \overline{S}_b^2/\sigma^2)\right)+
\frac{1}{2} \log\left(1+2\frac{\overline{S}_b}{\sigma} \frac{u_b^T
\varepsilon_b}{\parallel \varepsilon_b \parallel^2
+ \overline{S}_b^2/\sigma^2} \right) \nonumber\\
&=&  -\log\left(\frac{S_0}{\sigma}\right) +
\frac{1}{2}\log(\parallel \varepsilon_b \parallel^2 +
\overline{S}_b^2/\sigma^2) +
\frac{\overline{S}_b}{\sigma}\frac{\parallel \varepsilon_b
\parallel}{\parallel \varepsilon_b \parallel^2 +
\overline{S}_b^2/\sigma^2} \frac{u_b^T
\varepsilon_b}{\parallel \varepsilon_b \parallel} \nonumber\\
&& ~~~~~~~ + O_P\left(\frac{\overline{S}_b^2}{\sigma^2}
\frac{\parallel \varepsilon_b \parallel^2}{(\parallel \varepsilon_b
\parallel^2 + \overline{S}_b^2/\sigma^2)^2}\right).
\end{eqnarray}
Under the Rician noise mode (\ref{eq:S_b_def}),  $\varepsilon_b \sim
N(0,I_2)$, and so $\parallel \varepsilon_b\parallel^2$ has the
Exponential distribution with mean 2. Moreover, the third term on
the RHS of (\ref{eq:log_S_b_expand_small_S_b}) has mean 0. Thus this
expansion clearly indicates that under the regime where
$\overline{S}_b \ll \sigma$, the linear regression estimator
$\widehat{D}_{LS}$ will be biased .

To make the argument more concrete, we consider the scenario where
the gradient set ${\cal B}$ can be partitioned into ${\cal B}_0 \cup
{\cal B}_0^c$ such that, for $b \in {\cal B}_0$,
$\overline{S}_b^{-1} = o(\sigma^{-1})$ and for $b \in {\cal B}_0^c$,
$\overline{S}_b = o(\sigma)$. Then the linear regression estimator
$\widehat D_{LS}$ can be expressed as follows. Define
$\overline{S}^* := \min_{b \in {\cal B}} \overline{S}_b$,
$\overline{S}_* := \max_{b \in {\cal B}_0^c} \overline{S}_b$ and
$\overline{S}_{**} := \min_{b \in {\cal B}_0} \overline{S}_b$. Then,
\begin{eqnarray*}
\widehat D_{LS} &=& - (\sum_{b \in {\cal B}} x_b x_b^T)^{-1}
(\sum_{b \in {\cal B}} (\log S_b - \log S_0) x_b ) \nonumber\\
&=& (\sum_{b \in {\cal B}} x_b x_b^T)^{-1} \left[(\sum_{b \in {\cal
B}_0} x_b x_b^T) D^0 - (\sum_{b \in {\cal B}_0}
\frac{\sigma}{\overline{S}_b}
(u_b^T \varepsilon_b) x_b)\right] \nonumber\\
&& +(\sum_{b \in {\cal B}} x_b x_b^T)^{-1}\left[
\log\left(\frac{S_0}{\sigma}\right)\sum_{b \in {\cal B}_0^c} x_b  -
\frac{1}{2} \sum_{b \in {\cal B}_0^c} \log(\parallel
\varepsilon_b\parallel^2 + \overline{S}_b^2/\sigma^2)  x_b \right. \nonumber\\
&&  ~~~~ \left. - \sum_{b \in {\cal B}_0^c}
\frac{\overline{S}_b}{\sigma} \frac{\parallel \varepsilon_b
\parallel}{\parallel \varepsilon_b \parallel^2 +
\overline{S}_b^2/\sigma^2} \frac{u_b^T \varepsilon_b}{\parallel
\varepsilon_b \parallel}
x_b\right] \nonumber\\
&& ~~~~ +
O_P\left(\max\{\left(\frac{\sigma}{\overline{S}^*}\right)^2,
\left(\frac{\overline{S}_{*}}{\sigma}\right)^2\max_{b \in {\cal
B}_0^c} \frac{\parallel \varepsilon_b\parallel^2}{(\parallel
\varepsilon_b\parallel^2 +
(\overline{S}_{**}/\sigma)^2)^2}\}\right).
\end{eqnarray*}
The above expression can be decomposed as the summation of the bias
and variance terms up to the first order, which is more easily
interpretable:
\begin{eqnarray}\label{eq:D_hat_LS_small_signal}
\widehat D_{LS} - D^0  &=& (\sum_{b \in {\cal B}} x_b
x_b^T)^{-1}\left[-(\sum_{b \in {\cal B}_0^c} x_b x_b^T) D^0  +
\sum_{b \in {\cal B}_0^c} (\log(\frac{S_0}{\sigma}) -
\frac{1}{2}\log(\parallel \varepsilon_b\parallel^2 +
\overline{S}_b^2/\sigma^2)) x_b\right]
\nonumber\\
&& - (\sum_{b \in {\cal B}} x_b x_b^T)^{-1} \left[\sum_{b \in {\cal
B}_0} \frac{\sigma}{\overline{S}_b} (u_b^T \varepsilon_b) x_b +
\sum_{b \in {\cal B}_0^c} \frac{\overline{S}_b}{\sigma}
\frac{\parallel \varepsilon_b \parallel}{\parallel \varepsilon_b
\parallel^2 + \overline{S}_b^2/\sigma^2} \frac{u_b^T
\varepsilon_b}{\parallel \varepsilon_b\parallel} x_b \right]
\nonumber\\
&& ~~~~ +
O_P\left(\max\{\left(\frac{\sigma}{\overline{S}^*}\right)^2,
\left(\frac{\overline{S}_{*}}{\sigma}\right)^2\max_{b \in {\cal
B}_0^c} \frac{\parallel \varepsilon_b\parallel^2}{(\parallel
\varepsilon_b\parallel^2 +
(\overline{S}_{**}/\sigma)^2)^2}\}\right).
\end{eqnarray}
The second term on the RHS has mean zero and has the major
contribution in the variability of $\widehat D_{LS}$, while the
first term primarily contributes to its bias.

\bigskip

\subsection*{Acknowledgement}

Chen, Paul and Peng's research was partially supported by the NSF
grant DMS-0806128.

\section*{Appendix}

\subsection*{Perturbation analysis of tensor means}

\subsubsection*{Proof of Proposition \ref{prop:E_vs_LE}}

First consider case (a): the eigenvalues of $\bar{S}$ are distinct.
Let $\mu_j(\omega)$ denote the $j$-th largest eigenvalue of
$S(\omega)$ and $Q_j(\omega)$ denote the corresponding
eigen-projection. Then under (\ref{eq:eigen_perturb_cond}), for $t$
sufficiently small, with probability 1, $\mu_j(\omega)$ is of
multiplicity 1, and so $Q_j(\omega)$ is a rank 1 matrix. We can then
use the following matrix perturbation analysis results:
\begin{equation}\label{eq:eigenvalue_perturbation}
\mu_j(\omega) = \lambda_j + \tr(P_j B(\omega)) - \tr(P_j B(\omega)
H_j B(\omega)) + O(C^2 \parallel B \parallel_\infty^3),
\end{equation}
where $\tr$ denotes the trace of a matrix, and
\begin{eqnarray}\label{eq:eigenvector_perturbation}
Q_j(\omega) &=& P_j - (P_j B(\omega) H_j + H_j B(\omega)
P_j) \nonumber\\
&& + P_j B(\omega) H_j B(\omega) H_j + H_j B(\omega) P_j B(\omega)
H_j + H_j B(\omega) H_j B(\omega) P_j \nonumber\\
&& - P_j B(\omega) P_j B(\omega) H_j^2 - P_j B(\omega) H_j^2
B(\omega) P_j - H_j^2 B(\omega) P_j B(\omega) P_j  + O(C^3
\parallel B \parallel_\infty^3),
\end{eqnarray}
where $\parallel B \parallel_\infty := \sup_{\omega \in \Omega}
\parallel B(\omega) \parallel$.

In order to compare the Euclidean mean with the mean defined in
terms of the log-Euclidean metric, we consider an asymptotic
expansion of $\log S(\omega)$ around $\log \bar{S}$, where the
asymptotics is as $t \to 0$.

By definition, and using (\ref{eq:eigen_perturb_cond}), for $t$
sufficiently small,
\begin{eqnarray}\label{eq:log_S_eigen_expansion}
&& \log S(\omega) - \log \bar{S} \nonumber\\
&=& \sum_{j=1}^N (\log\mu_j(\omega)
Q_j(\omega) - \log \lambda_j P_j) \nonumber\\
&=& \sum_{j=1}^N (\log\mu_j(\omega) - \log \lambda_j) P_j +
\sum_{j=1}^N \log \lambda_j (Q_j(\omega) - P_j)  + \sum_{j=1}^N
(\log\mu_j(\omega) - \log \lambda_j) (Q_j(\omega) - P_j) \nonumber\\
&&
\end{eqnarray}
Now, from (\ref{eq:eigenvalue_perturbation}) and
(\ref{eq:eigen_perturb_cond}), we have
\begin{eqnarray*}
\log \mu_j(\omega) - \log \lambda_j &=& \log\left(1 +
\frac{\mu_j(\omega) - \lambda_j}{\lambda_j}\right) \\
&=& \frac{1}{\lambda_j} [\tr(P_j B(\omega))- \tr(P_j B(\omega) H_j
B(\omega))] - \frac{1}{2\lambda_j^2} [\tr(P_j B(\omega))]^2 +
O(t^3),
\end{eqnarray*}
where we have used the series expansion $\log(1+x) =
\sum_{n=1}^\infty (-1)^{n-1} x^n/n$. Substituting in
(\ref{eq:log_S_eigen_expansion}) and using
(\ref{eq:eigenvector_perturbation}), we get,
\begin{eqnarray*}
&& \log S - \log \bar{S} \nonumber\\
&=& \sum_{j=1}^N \frac{1}{\lambda_j} \tr(P_j B(\omega)) P_j -
\sum_{j=1}^N \frac{1}{\lambda_j} \tr(P_j B(\omega) H_j B(\omega))
P_j - \frac{1}{2} \sum_{j=1}^N \frac{1}{\lambda_j^2} [\tr(P_j
B(\omega))]^2 P_j \nonumber\\
&& - \sum_{j=1}^N \log \lambda_j [P_j B(\omega) H_j + H_j B(\omega)
P_j] \nonumber\\
&& + \sum_{j=1}^N \log \lambda_j \left[ P_j B(\omega) H_j B(\omega)
H_j + H_j B(\omega) P_j B(\omega)
H_j + H_j B(\omega) H_j B(\omega) P_j \right. \nonumber\\
&& \left. - P_j B(\omega) P_j B(\omega) H_j^2 - P_j B(\omega) H_j^2
B(\omega) P_j - H_j^2 B(\omega) P_j B(\omega) P_j\right] \nonumber\\
&& - \sum_{j=1}^N \frac{1}{\lambda_j} \tr(P_jB(\omega)) (P_j
B(\omega) H_j + H_j B(\omega) P_j) + O(t^3).
\end{eqnarray*}
From this, we get (\ref{eq:log_S_mean_expansion}) by taking
expectation, and recalling that $\mathbb{E}(B) = O$.

Now consider case (b): $\bar{S} = \lambda_1 I$. Then, from the
expansion
\begin{eqnarray*}
\log S(\omega)  &=& (\log\lambda_1) I + \log\left( I +
\frac{1}{\lambda_1}B(\omega)\right) ~=~ \log \bar{S} +
\frac{1}{\lambda_1}B(\omega) - \frac{1}{2\lambda_1^2}(B(\omega))^2 +
O(C^3 \parallel B
\parallel_\infty^3),
\end{eqnarray*}
we obtain (\ref{eq:log_S_mean_expansion_isotropic}) by taking
expectation.

\subsubsection*{Proof of Proposition \ref{prop:E_vs_Affine}}

Define,
\begin{equation*}
L(\omega) = \log(\bar{S}_{Aff}^{-1/2} S(\omega)
\bar{S}_{Aff}^{-1/2}), \qquad \omega \in \Omega.
\end{equation*}
Note that, (\ref{eq:affine_S_mean_barycentric}) implies that
$\mathbb{E}(L) = \mathbf{0}$. First, we show that
\begin{equation}\label{eq:L_sup_order}
\parallel L \parallel_\infty := \sup_{\omega \in \Omega} \parallel L \parallel =
O(t).
\end{equation}
This implies that
\begin{eqnarray*}
\bar{S} &=& \mathbb{E}(S) ~=~ \mathbb{E}(\bar{S}_{Aff}^{1/2}e^L
\bar{S}_{Aff}^{1/2}) \nonumber\\
&=& \bar{S}_{Aff} + \bar{S}_{Aff}^{1/2} \mathbb{E}(L)
\bar{S}_{Aff}^{1/2} + \frac{1}{2} \bar{S}_{Aff}^{1/2}
\mathbb{E}(L^2) \bar{S}_{Aff}^{1/2} + O(t^3),
\end{eqnarray*}
from which we get
\begin{equation}\label{eq:S_compare_mean_Euclidean_affine}
\bar{S}_{Aff} = \bar{S} - \frac{1}{2} \bar{S}^{1/2} \mathbb{E}(L^2)
\bar{S}^{1/2} + O(t^3),
\end{equation}
where, we have used the fact that $\mathbb{E}(L) = \mathbf{0}$. Now,
we express $\mathbb{E}(L^2)$ in terms of an expectation involving $B
= S - \bar{S}$. In order to do this, observe that
\begin{eqnarray*}
B(\omega) &=& S(\omega) - \bar{S} ~=~ \bar{S}_{Aff}^{1/2}
e^{L(\omega)} \bar{S}_{Aff}^{1/2} - \bar{S}_{Aff} + \bar{S}_{Aff} -
\bar{S} \\
&=& \bar{S}_{Aff}^{1/2} L(\omega) \bar{S}_{Aff}^{1/2} + \frac{1}{2}
\bar{S}_{Aff}^{1/2}( (L(\omega))^2 - \mathbb{E}(L^2))
\bar{S}_{Aff}^{1/2} + O(t^3)\\
&=& \bar{S}^{1/2} L(\omega) \bar{S}^{1/2} + \frac{1}{2}
\bar{S}^{1/2}( (L(\omega))^2 - \mathbb{E}(L^2)) \bar{S}^{1/2} +
O(t^3),
\end{eqnarray*}
where, in the second and third steps we have used
(\ref{eq:L_sup_order}) and
(\ref{eq:S_compare_mean_Euclidean_affine}). Therefore, again using
(\ref{eq:L_sup_order}),
\begin{equation}\label{eq:E_L_square}
\mathbb{E}(L^2) = \bar{S}^{-1/2}\mathbb{E}(B \bar{S}^{-1} B)
\bar{S}^{-1/2} + O(t^3),
\end{equation}
which, together with (\ref{eq:S_compare_mean_Euclidean_affine}),
leads to the representation
\begin{equation}\label{eq:S_compare_mean_Euclidean_affine_B}
\bar{S}_{Aff} = \bar{S} - \frac{1}{2} \mathbb{E}(B \bar{S}^{-1} B) +
O(t^3).
\end{equation}
Now we are in a position to complete the proof. For case (a), using
similar perturbation analysis as in the expansion of $\log
S(\omega)$ around $\log \bar{S}$, and then using
(\ref{eq:S_compare_mean_Euclidean_affine_B}), which in particular
implies that $\parallel \bar{S}_{Aff} - \bar{S} \parallel = O(t^2)$,
we obtain (\ref{eq:log_S_affine_mean_expansion}).

For case (b), since $\bar{S} = \lambda_1 I$, from
(\ref{eq:S_compare_mean_Euclidean_affine_B}), and using series
expansion of $\log(I + K)$, where $K$ is symmetric,
(\ref{eq:log_S_affine_mean_expansion_isotropic}) immediately
follows.

\subsubsection*{Proof of (\ref{eq:L_sup_order})}

From the fact that $\parallel B \parallel_\infty = O(t)$, it easily
follows that $\mathbb{E}(d_{Aff}^2(\bar{S},S)) = O(t^2)$. Then, by
definition of $\bar{S}_{Aff}$ it follows that
$\mathbb{E}(d_{Aff}^2(\bar{S}_{Aff},S)) = O(t^2)$ which implies
$\mathbb{E}(d_{Aff}(\bar{S}_{Aff},S)) = O(t)$. Now, writing
\begin{equation*}
\log(\bar{S}_{Aff}^{-1/2} S(\omega) \bar{S}_{Aff}^{-1/2}) =
\log(\bar{S}_{Aff}^{-1/2} \bar{S} \bar{S}_{Aff}^{-1/2} +
\bar{S}_{Aff}^{-1/2} B(\omega) \bar{S}_{Aff}^{-1/2})
\end{equation*}
and using the Baker-Campbell-Hausdorff formula, together with the
fact that $\parallel B \parallel_\infty = O(t)$, we conclude that
\begin{equation*}
d_{Aff}(\bar{S}_{Aff},\bar{S}) = O(t) ~~\mbox{so that}~~ \parallel
\bar{S}_{Aff} - \bar{S} \parallel = O(t).
\end{equation*}
From this it is easy to deduce (\ref{eq:L_sup_order}).

\subsection*{Rician noise model}

\subsubsection*{Facts about Bessel functions}

The following result about Bessel functions (cf. Abramowitz and
Stegun, 1965) is crucial in proving results involving the Rician
noise model.
\begin{lemma}\label{lemma:Bessel_derivative}
For every $\nu$, and $t \neq 0$
\begin{equation}\label{eq:Bessel_function_facts}
I_{\nu+1}(t) = I_{\nu-1}(t) - \frac{2\nu}{t} I_\nu(t) \qquad
\mbox{and} \qquad I_\nu'(t) = I_{\nu-1}(t) - \frac{\nu}{t} I_\nu(t)
= I_{\nu+1}(t) + \frac{\nu}{t} I_{\nu}(t),
\end{equation}
where $I_\nu'(t)$ denotes the derivative of $I_\nu(t)$ with respect
to $t$.
\end{lemma}

We list down some facts about the function
\begin{equation}\label{eq:I_1_I_0_ratio}
F(t) := \frac{I_1(t)}{I_0(t)}, \qquad t \in \mathbb{R}
\end{equation}
that we have used in deriving the asymptotic expansion for the MLE
of $D$ in the limit $\sigma \to 0$.
\begin{enumerate}
\item
Using (\ref{eq:Bessel_function_facts}), we have a differential
equation for $F$,
\begin{equation}\label{eq:I_1_I_0_ratio_ODE}
F'(t) = 1 - \frac{1}{t} F(t) - (F(t))^2 ~~~\mbox{for}~~~t \neq 0.
\end{equation}
\item
$F(t) \to 1$ and $t(1-F(t)) \to 1/2$ as $t \to \infty$.
\item
Above facts, together with (\ref{eq:I_1_I_0_ratio_ODE}) imply that
$tF'(t) \to 0$ and $tF''(t) \to 0$ as $t \to \infty$.
\end{enumerate}

\vskip.1in\noindent{\bf Proof of Proposition
\ref{prop:Fisher_Rician} :} Let $\nabla \log
p_{\zeta(D,b),\sigma}(x)$ denote the gradient of $\log
p_{\zeta(D,b),\sigma}(x)$ with respect to the parameter $D$.  Using
the chain rule of differentiation, we have
\begin{equation}
\nabla \log p_{\zeta(D,b),\sigma}(x) = \frac{\partial}{\partial
\zeta} \log p_{\zeta,\sigma}(x)\left|_{\zeta = \zeta(D,b)}\right.
\nabla_D \zeta(D,b) = \frac{\partial}{\partial \zeta} \log
p_{\zeta,\sigma}(x)\left|_{\zeta = \zeta(D,b)}\right.  \cdot
(-\zeta(D,b)) x_b.
\end{equation}
It is enough therefore to find the Fisher information for $\zeta$
under the Rician noise model with parameter $(\zeta,\sigma)$ (with
$\sigma$ known). Now, it follows from
(\ref{eq:Bessel_function_facts}) that $I_0'(t) = I_1(t)$. Thus,
\begin{eqnarray}
\frac{\partial}{\partial \zeta}\log p_{\zeta,\sigma}(x) &=& -
\frac{\zeta}{\sigma^2} + \frac{x}{\sigma^2}
\frac{I_1\left(\frac{x\zeta}{\sigma^2}\right)}{I_0\left(\frac{x\zeta}{\sigma^2}\right)}.
\label{eq:zeta_score}
\end{eqnarray}
Since $\mathbb{E}(\frac{\partial}{\partial \zeta}\log
p_{\zeta,\sigma}(X)) = 0$, from (\ref{eq:zeta_score}) it follows
that the Fisher information for $\zeta$ is given by,
\begin{eqnarray}\label{eq:zeta_score_var}
\mathfrak{I}(\zeta;\sigma) := \mbox{Var}(\frac{\partial}{\partial
\zeta}\log p_{\zeta,\sigma}(X)) &=&
\mathbb{E}\left[\frac{X}{\sigma^2}
\frac{I_1\left(\frac{X\zeta}{\sigma^2}\right)}{I_0\left(\frac{X\zeta}{\sigma^2}\right)}\right]^2
- \frac{\zeta^2}{\sigma^4}\nonumber\\
&=& e^{-\frac{\zeta^2}{2\sigma^2}}\int_0^\infty
\left(\frac{x}{\sigma^2}\right)^3
\frac{(I_1\left(\frac{x\zeta}{\sigma^2}\right))^2}{I_0\left(\frac{x\zeta}{\sigma^2}\right)}
e^{-\frac{x^2}{2\sigma^2}} dx - \frac{\zeta^2}{\sigma^4}\nonumber\\
&=& \frac{1}{\sigma^2}\left[e^{-\frac{\zeta^2}{2\sigma^2}}
\int_0^\infty u^3 \frac{(I_1(u\zeta/\sigma))^2}{I_0(u\zeta/\sigma)}
e^{-u^2/2} du -  \frac{\zeta^2}{\sigma^2}\right] \nonumber\\
&=& \frac{1}{\sigma^2}\left[ e^{-\frac{\zeta^2}{2\sigma^2}}
V\left(\frac{\zeta}{\sigma}\right)- \frac{\zeta^2}{\sigma^2}\right],
\end{eqnarray}
where, $V(w)$ is given by (\ref{eq:V_w_expr}).

\subsubsection*{Outline of asymptotic expansions of $\widehat D_{LS}$ and $\widehat D_{NL}$}

Define, for an arbitrary tensor $D$ and arbitrary gradient direction
$b$,
\begin{equation}\label{eq:f_b_D}
f_b(w,D) := \parallel S_0 e^{-x_b^T D} u_b +  w\parallel.
\end{equation}
Then, $f_b(0,D^0) = \overline{S}_b$ and $f_b(\sigma
\varepsilon_b,D^0) = S_b$. We denote the partial derivatives
with respect to the first and second arguments by $\nabla_{ij} f_b$
etc, where $1\leq i,j\leq 2$. Then,
\begin{equation}\label{eq:grad_1_f_b}
\nabla_1 f_b(w,D^0) = \frac{1}{f_b(w,D^0)} (\overline{S}_b u_b + w)
~~\Longrightarrow~~ \nabla_1 f_b(0,D^0) = \frac{1}{\overline{S}_b}
\overline{S}_b u_b = u_b.
\end{equation}
Now, let $v_b \in \mathbb{R}^2$ be such that $v_b^T u_b = 0$ and
$\parallel v_b \parallel =1$ so that $u_b u_b^T + v_b v_b^T = I_2$.
Thus, using (\ref{eq:grad_1_f_b}),
\begin{eqnarray}\label{eq:hess_1_f_b}
\nabla_{11} f_b(w,D^0) &=& \frac{1}{f_b(w,D^0)} I_2 -
\frac{1}{(f_b(w,D^0))^3}
(\overline{S}_b u_b + w)(\overline{S}_b u_b + w)^T  \nonumber\\
&\Longrightarrow &  \nabla_{11} f_b(0,D^0) =
\frac{1}{\overline{S}_b} (I_2 - u_bu_b^T) = \frac{1}{\overline{S}_b}
v_b v_b^T.
\end{eqnarray}
We also have,
\begin{equation}\label{eq:grad_2_f_b}
\nabla_2 f_b(0,D^0) = - S_0 \exp(-x_b^T D^0) x_b = - \overline{S}_b
x_b
\end{equation}
and
\begin{equation}\label{eq:hess_2_f_b}
\nabla_{22} f_b(0,D^0) = S_0 \exp(-x_b^T D^0) x_bx_b^T =
\overline{S}_b x_bx_b^T.
\end{equation}

\vskip.1in\noindent{\bf Proof of Proposition
\ref{prop:D_hat_LS_asymp} :}
\begin{eqnarray*}
\widehat D_{LS} &=& - (\sum_{b \in {\cal B}} x_b x_b^T)^{-1}(\sum_{b
\in {\cal B}} (\log S_b - \log S_0) x_b) \nonumber\\
&=&  - (\sum_{b \in {\cal B}} x_b x_b^T)^{-1}(\sum_{b
\in {\cal B}}  (\log f_b(\sigma \varepsilon_b,D^0) - \log f_b(0,D^0) - x_b^T D^0) x_b ) \nonumber\\
&=&  D^0 - (\sum_{b \in {\cal B}} x_b x_b^T)^{-1} \left(\sum_{b \in
B}
\sigma \frac{(\nabla_1 f_b(0,D^0))^T \varepsilon_b}{f_b(0,D^0)} x_b\right) \nonumber\\
&& - \frac{1}{2}(\sum_{b \in {\cal B}} x_b x_b^T)^{-1} \sum_{b \in
{\cal B}} \sigma^2 \left[  \frac{\varepsilon_b^T \nabla_{11}
f_b(0,D^0) \varepsilon_b}{f_b(0,D^0)} - \frac{(\varepsilon_b^T
\nabla_{1} f_b(0,D^0))^2}{(f_b(0,D^0))^2}\right] x_b +
O_P(\sigma^3).
\end{eqnarray*}
This, together with (\ref{eq:grad_1_f_b}) and (\ref{eq:hess_1_f_b}),
proves Proposition \ref{prop:D_hat_LS_asymp}.

\vskip.1in\noindent{\bf Proof of Proposition
\ref{prop:D_hat_NL_asymp} :} From this (\ref{eq:NL_normal}), and the
definition of $\widehat D_{NL}$, using standard arguments, it can be
shown that $\parallel \widehat D_{NL} - D^0
\parallel = O_P(\sigma)$ as $\sigma \to 0$. Then, expanding the LHS
of (\ref{eq:NL_normal}) in Taylor series, we have
\begin{eqnarray*}
\sum_{b \in {\cal B}}  \left[\sigma \varepsilon_b^T \nabla_1
f_b(0,D^0) + \frac{\sigma^2}{2} \varepsilon_b^T \nabla_{11}
f_b(0,D^0) \varepsilon_b
- (\nabla_2 f_b(0,D^0))^T (\widehat D_{NL} - D^0) \right. && \\
\left. -\frac{1}{2}
(\widehat D_{NL} - D^0)^T \nabla_{22} f_b(0,D^0) (\widehat D_{NL} - D^0)\right] ~~~~~~~~~~&& \\
~~\cdot~ \left(\nabla_2 f_b(0,D^0) + \nabla_{22} f_b(0,D^0)
(\widehat D_{NL} - D^0)\right) &=& O_P(\sigma^3).
\end{eqnarray*}
Now, changing sides and collecting terms corresponding to a given
power of $\sigma$, we can express $\widehat D_{NL}$ as $\widehat
D_{NL} = D^0 + \sigma D_{1,NL} + \sigma^2 D_{2,NL} + O_P(\sigma^3)$,
where
\begin{eqnarray*}
D_{1,NL} &=& (\sum_{b \in {\cal B}} \nabla_2 f_b(0,D^0) (\nabla_2
f_b(0,D^0))^T)^{-1} (\sum_{b \in {\cal B}}  \varepsilon_b^T \nabla_1
f_b(0,D^0)   \nabla_2 f_b(0,D^0) )
\end{eqnarray*}
which equals (\ref{eq:D_1_NL}) by virtue of (\ref{eq:grad_2_f_b}).
Next, we have
\begin{eqnarray*}
D_{2,NL} &=& (\sum_{b \in {\cal B}} \nabla_2 f_b(0,D^0) (\nabla_2
f_b(0,D^0))^T)^{-1}
\cdot \nonumber\\
&& ~~\left[\frac{1}{2} \sum_{b \in {\cal B}} \left(\varepsilon_b^T
\nabla_{11} f_b(0,D^0) \varepsilon_b
- D_{1,NL}^T \nabla_{22}f_b(0,D^0) D_{1,NL}\right) \nabla_2 f_b(0,D^0) \right. \nonumber\\
&& + \left. \sum_{b \in {\cal B}}  (\varepsilon_b^T \nabla_1
f_b(0,D^0)) \nabla_{22} f_b(0,D^0) D_{1,NL} -  \sum_{b \in {\cal B}}
( D_{1,NL}^T \nabla_2 f_b(0,D^0)) \nabla_{22} f_b(0,D^0)
D_{1,NL}\right].
\end{eqnarray*}
Now, using (\ref{eq:grad_1_f_b})-(\ref{eq:hess_2_f_b}) and
(\ref{eq:D_1_NL}), we can simplify the expression for $D_{2,NL}$ as
\begin{eqnarray*}
D_{2,NL} &=& (\sum_{b \in {\cal B}} \overline{S}_b^2 x_b x_b^T)^{-1}
\left[\frac{3}{2} \sum_{b \in {\cal B}}  \overline{S}_b^2 (x_b^T
D_{1,NL})^2  x_b  - \frac{1}{2} \sum_{b \in {\cal B}} (v_b^T
\varepsilon_b)^2 x_b + \sum_{b\in {\cal B}} \overline{S}_b (u_b^T
\varepsilon_b) (x_b^T D_{1,NL}) x_b \right]
\end{eqnarray*}
which can be rearranged to get (\ref{eq:D_2_NL}).

\clearpage
\newpage

\subsection*{Figures for  Rician noise model}\label{subsec:plots_Rician}

\begin{figure}[th]\label{fig:median_error_Rician_sigma_0.1_L}
\begin{center}
\includegraphics[height=6.5in,width=5in,angle=270]{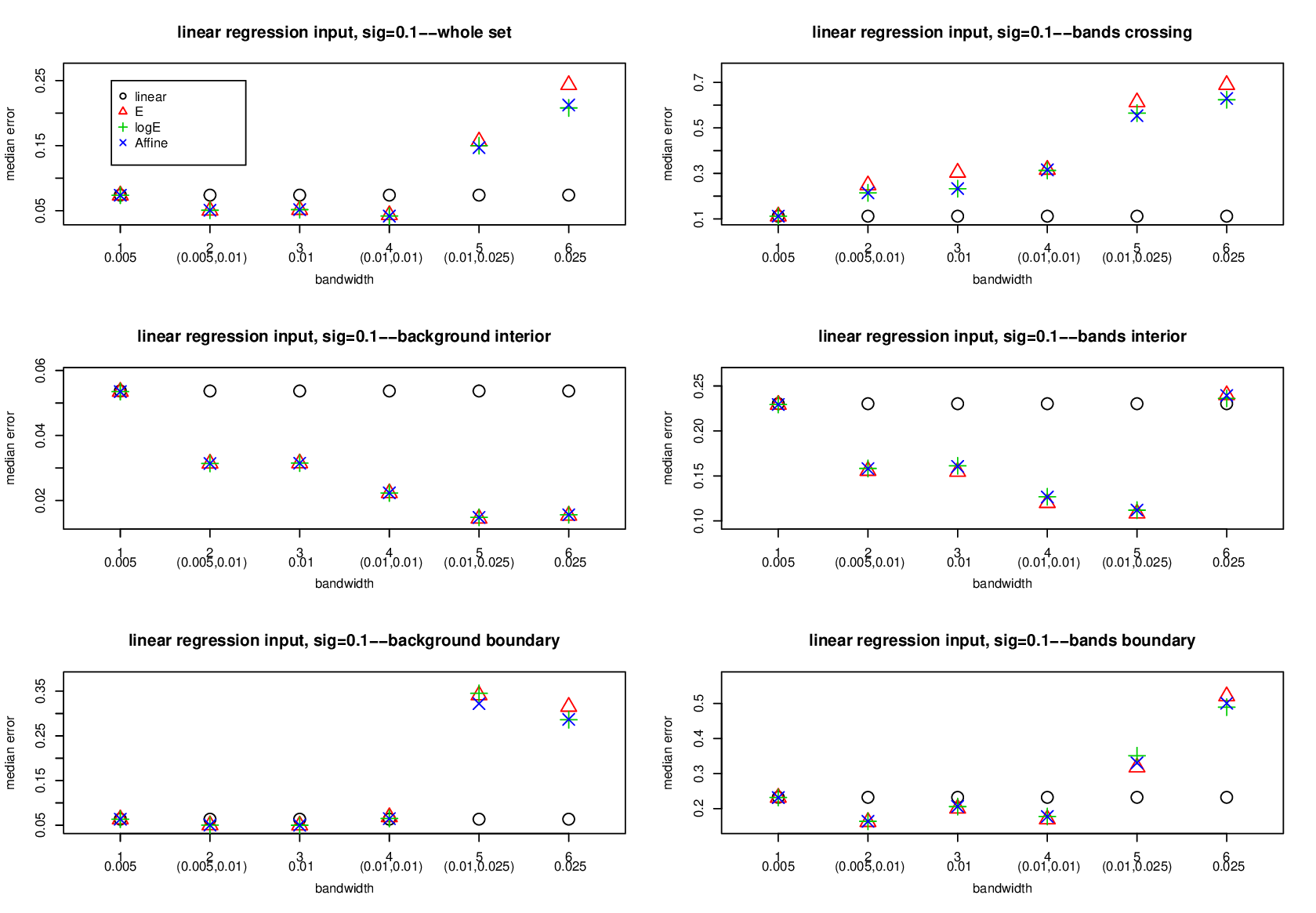}\\
\end{center}
\caption{{\bf Rician noise} with $\sigma =
0.1$. Comparison of median errors over different regions for
``observed'' tensors (obtained by {\bf linear regression}, represented
by black circle) and Euclidean (red triangle), log-Euclidean (green $+$) and
Affine smoothers (blue $\times$).}
\end{figure}

\begin{figure}[th]\label{fig:median_error_Rician_sigma_0.1_NL}
\begin{center}
\includegraphics[height=6.5in,width=5in,angle=270]{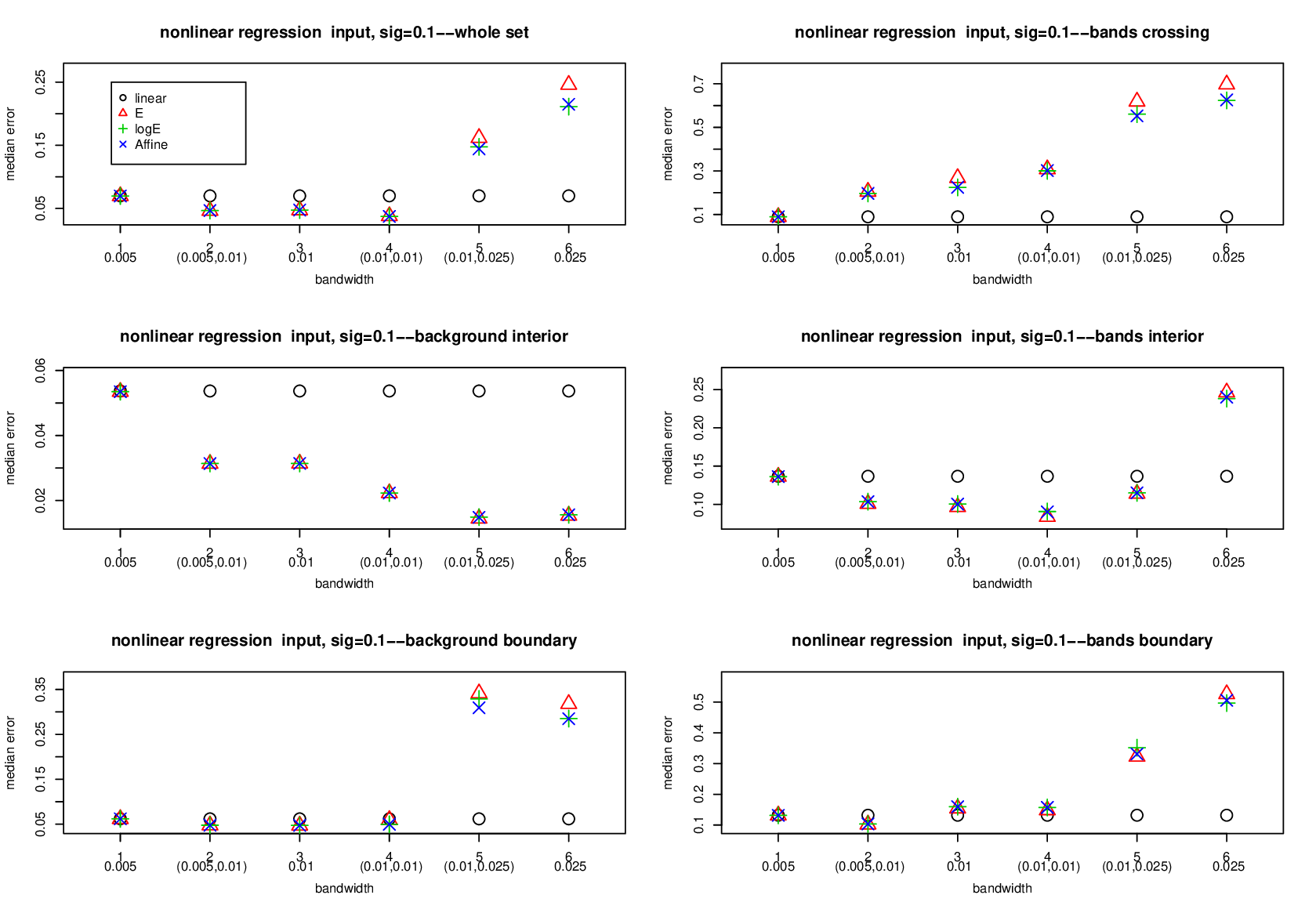}\\
\end{center}
\caption{{\bf Rician noise} with $\sigma =
0.1$. Comparison of median errors over different regions for
``observed'' tensors (obtained by {\bf nonlinear regression}, represented
by black circle) and Euclidean (red triangle), log-Euclidean (green $+$) and
Affine smoothers (blue $\times$).}
\end{figure}

\begin{figure}[th]\label{fig:median_error_Rician_sigma_0.5_L}
\begin{center}
\includegraphics[height=6.5in,width=5in,angle=270]{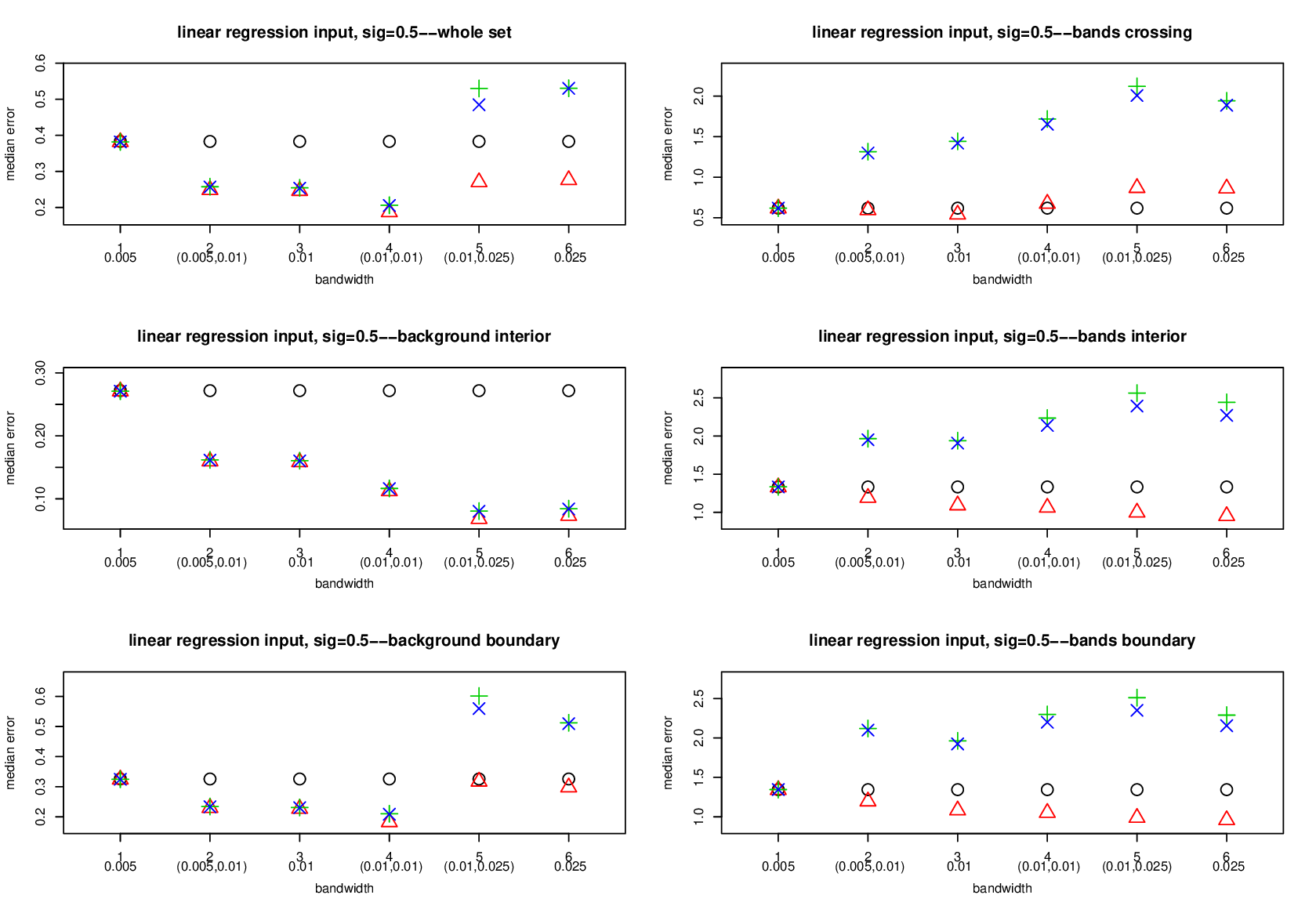}\\
\end{center}
\caption{{\bf Rician noise} with $\sigma =
0.5$. Comparison of median errors over different regions for
``observed'' tensors (obtained by {\bf linear regression}, represented
by black circle) and Euclidean (red triangle), log-Euclidean (green $+$) and
Affine smoothers (blue $\times$).}
\end{figure}

\begin{figure}[th]\label{fig:median_error_Rician_sigma_0.5_NL}
\begin{center}
\includegraphics[height=6.5in,width=5in,angle=270]{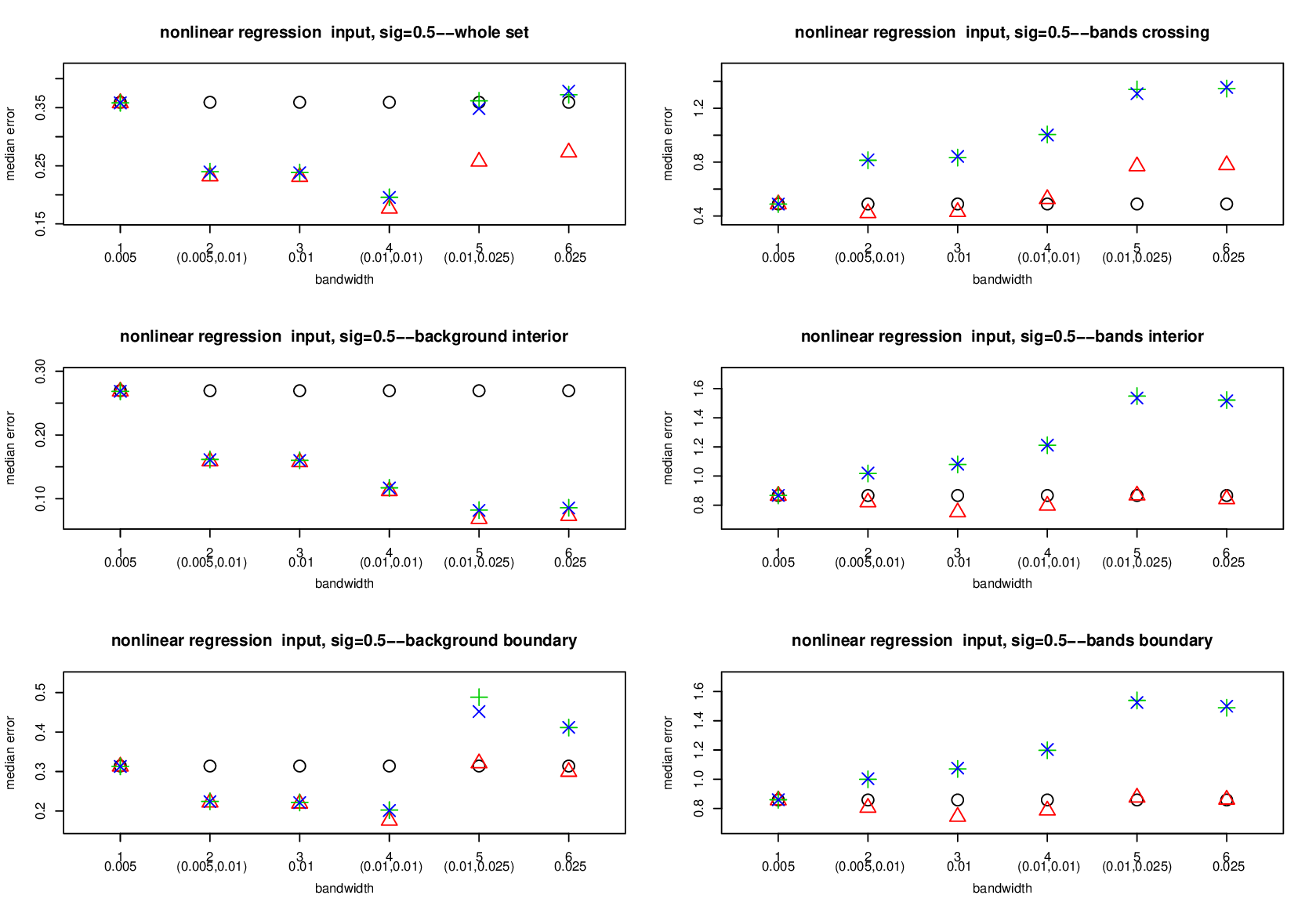}\\
\end{center}
\caption{{\bf Rician noise} with $\sigma = 0.5$.
Comparison of median errors over different regions for
``observed'' tensors (obtained by {\bf nonlinear regression}, represented
by black circle) and Euclidean (red triangle), log-Euclidean (green $+$) and
Affine smoothers (blue $\times$).}
\end{figure}

\begin{figure}[th]\label{fig:median_error_Rician_sigma_1_L}
\begin{center}
\includegraphics[height=6.5in,width=5in,angle=270]{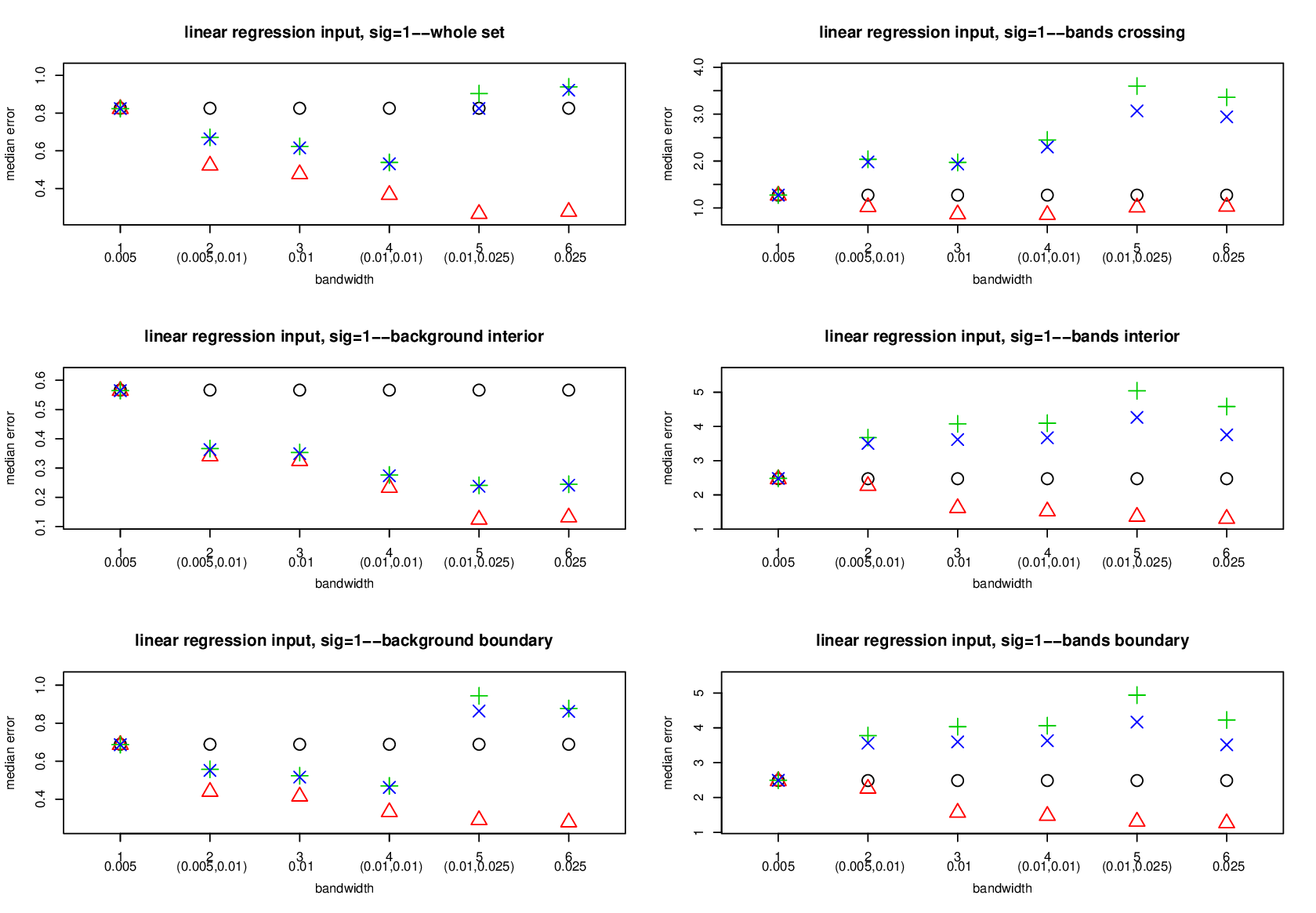}\\
\end{center}
\caption{{\bf Rician noise} with $\sigma =
1$. Comparison of median errors over different regions for
``observed'' tensors (obtained by {\bf linear regression}, represented
by black circle) and Euclidean (red triangle), log-Euclidean (green $+$) and
Affine smoothers (blue $\times$).}
\end{figure}

\begin{figure}[th]\label{fig:median_error_Rician_sigma_1_NL}
\begin{center}
\includegraphics[height=6.5in,width=5in,angle=270]{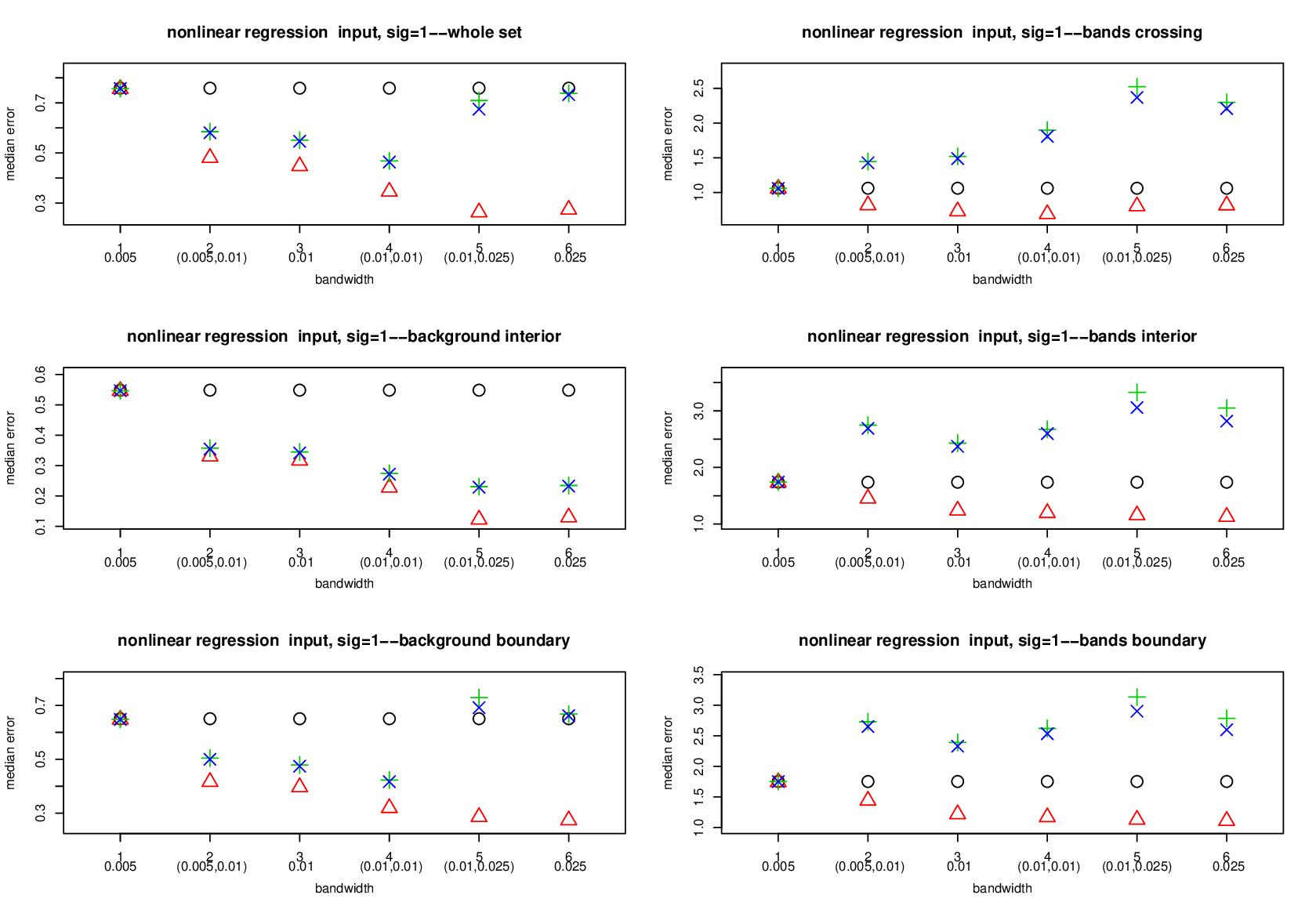}\\
\end{center}
\caption{{\bf Rician noise} with $\sigma =
1$. Comparison of median errors over different regions for
``observed'' tensors (obtained by {\bf nonlinear regression}, represented
by black circle) and Euclidean (red triangle), log-Euclidean (green $+$) and
Affine smoothers (blue $\times$).}
\end{figure}

\clearpage
\newpage

\subsection*{Figures for spectral noise}\label{subsec:plots_spectral}

\begin{figure}[th]\label{fig:median_error_spectral_dfeval_50_eta_0.1}
\begin{center}
\includegraphics[height=6.5in,width=5in,angle=270]{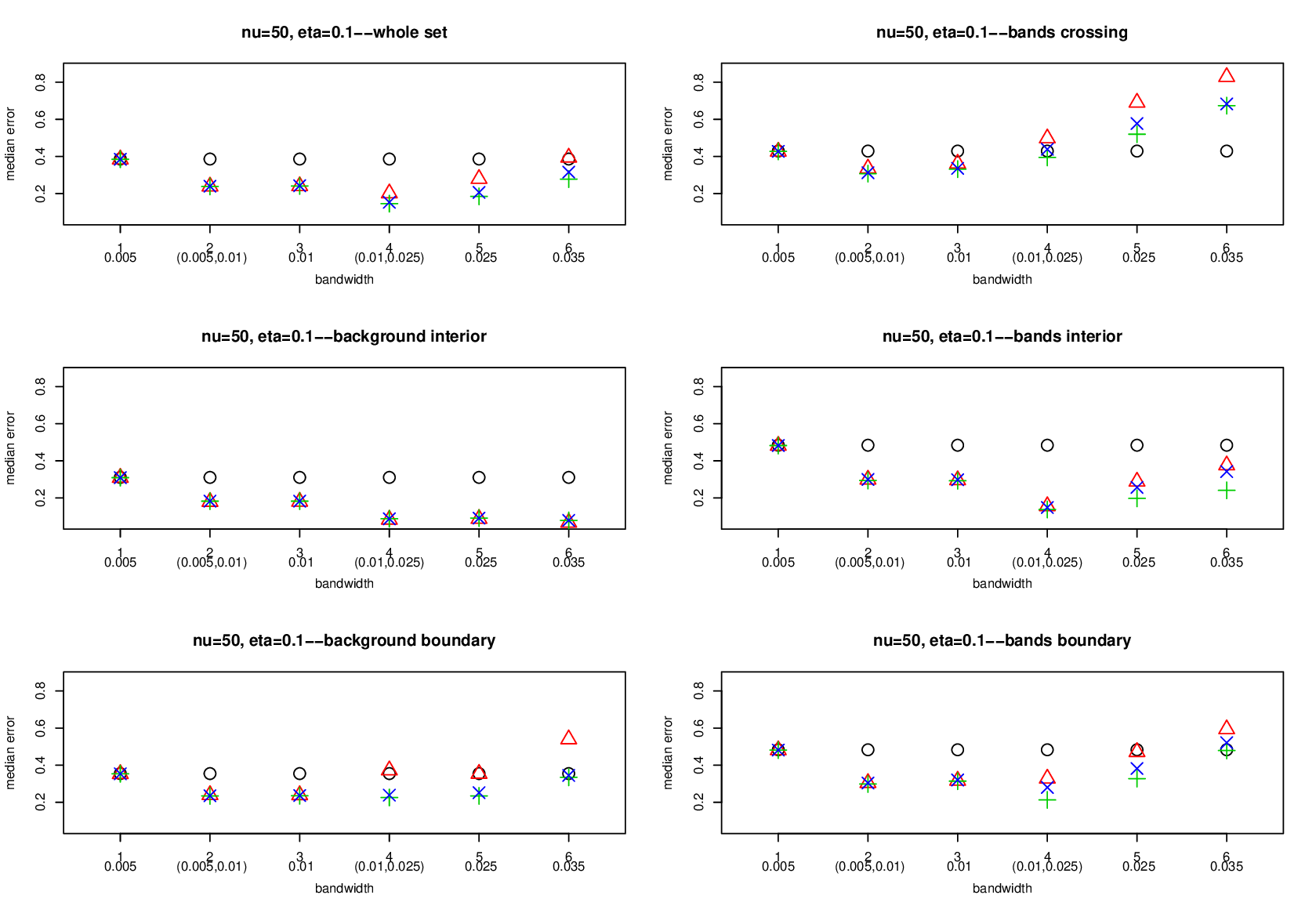}\\
\end{center}
\caption{{\bf Spectral noise} with $\nu = 50$
and $\eta=0.1$. Comparison of median errors over different regions for
observed tensors (black circle) and Euclidean (red triangle), log-Euclidean (green $+$) and
Affine smoothers (blue $\times$).}
\end{figure}

\begin{figure}[th]\label{fig:median_error_spectral_dfeval_50_eta_0.2}
\begin{center}
\includegraphics[height=6.5in,width=5in,angle=270]{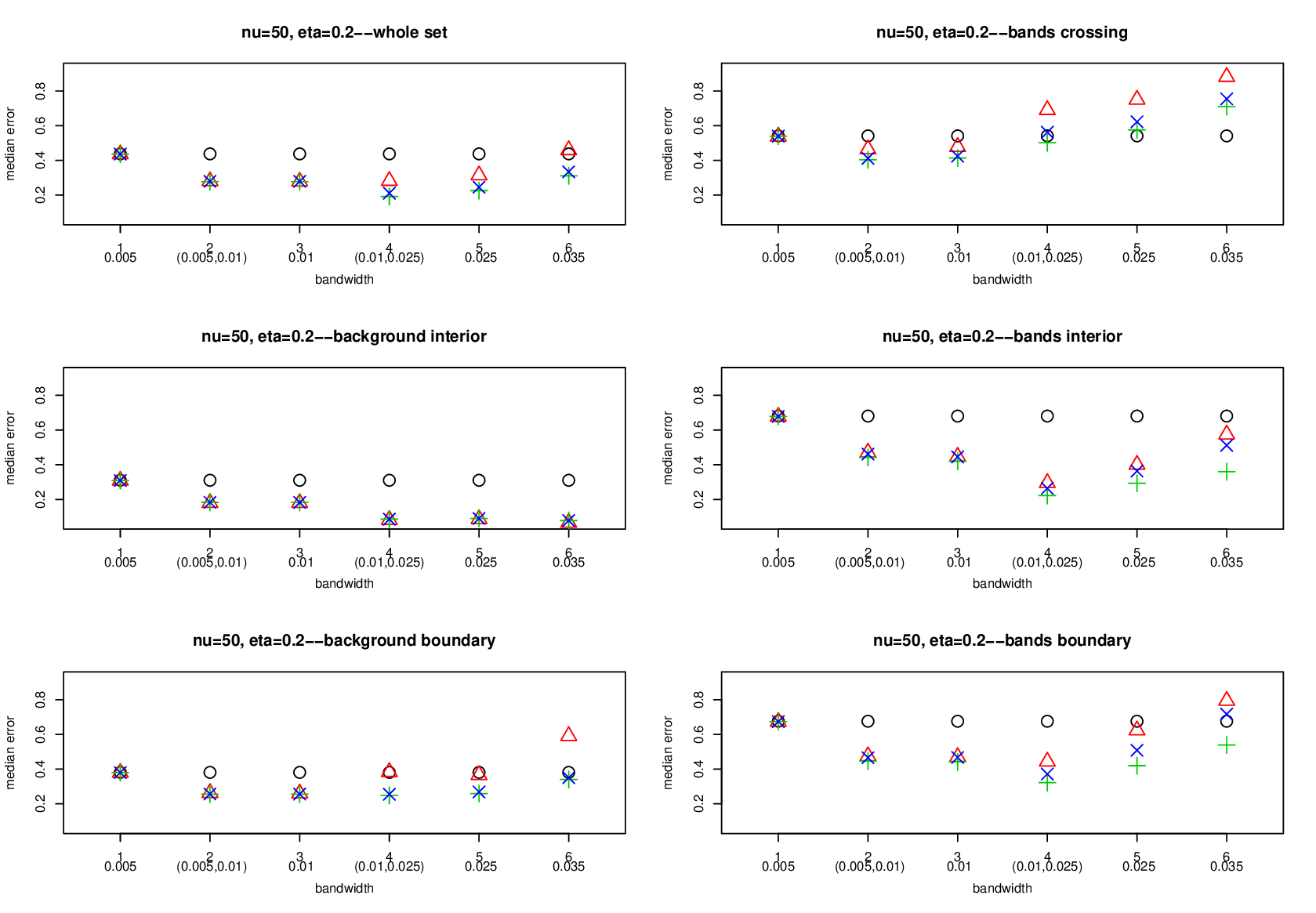}\\
\end{center}
\caption{{\bf Spectral noise} with $\nu = 50$
and $\eta=0.2$. Comparison of median errors over different regions for
observed tensors (black circle) and Euclidean (red triangle), log-Euclidean (green $+$) and
Affine smoothers (blue $\times$).}
\end{figure}

\begin{figure}[th]\label{fig:median_error_spectral_dfeval_50_eta_0.3}
\begin{center}
\includegraphics[height=6.5in,width=5in,angle=270]{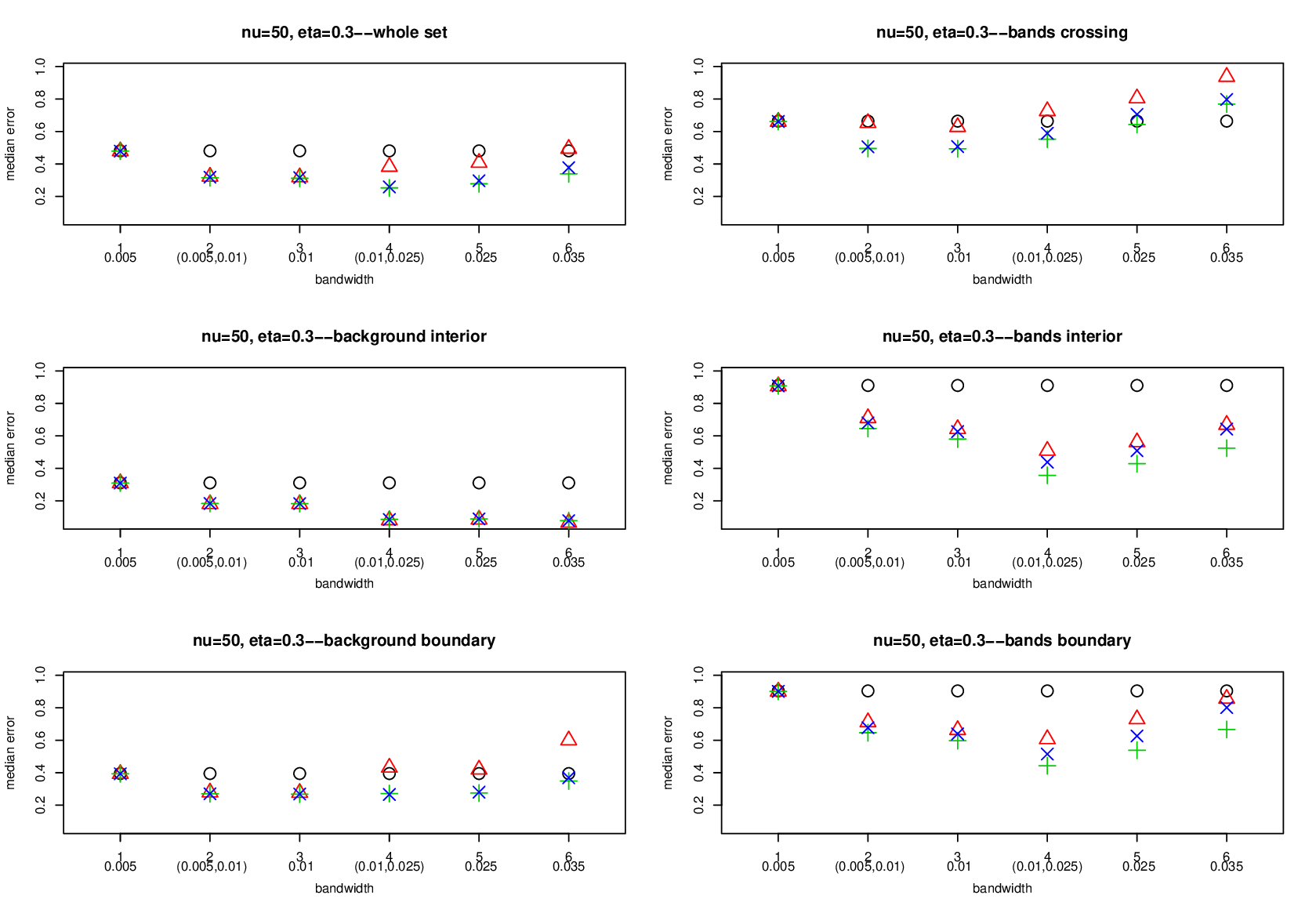}\\
\end{center}
\caption{{\bf Spectral noise} with $\nu = 50$
and $\eta=0.3$. Comparison of median errors over different regions for
observed tensors (black circle) and Euclidean (red triangle), log-Euclidean (green $+$) and
Affine smoothers (blue $\times$).}
\end{figure}

\begin{figure}[th]\label{fig:median_error_spectral_dfeval_20_eta_0.1}
\begin{center}
\includegraphics[height=6.5in,width=5in,angle=270]{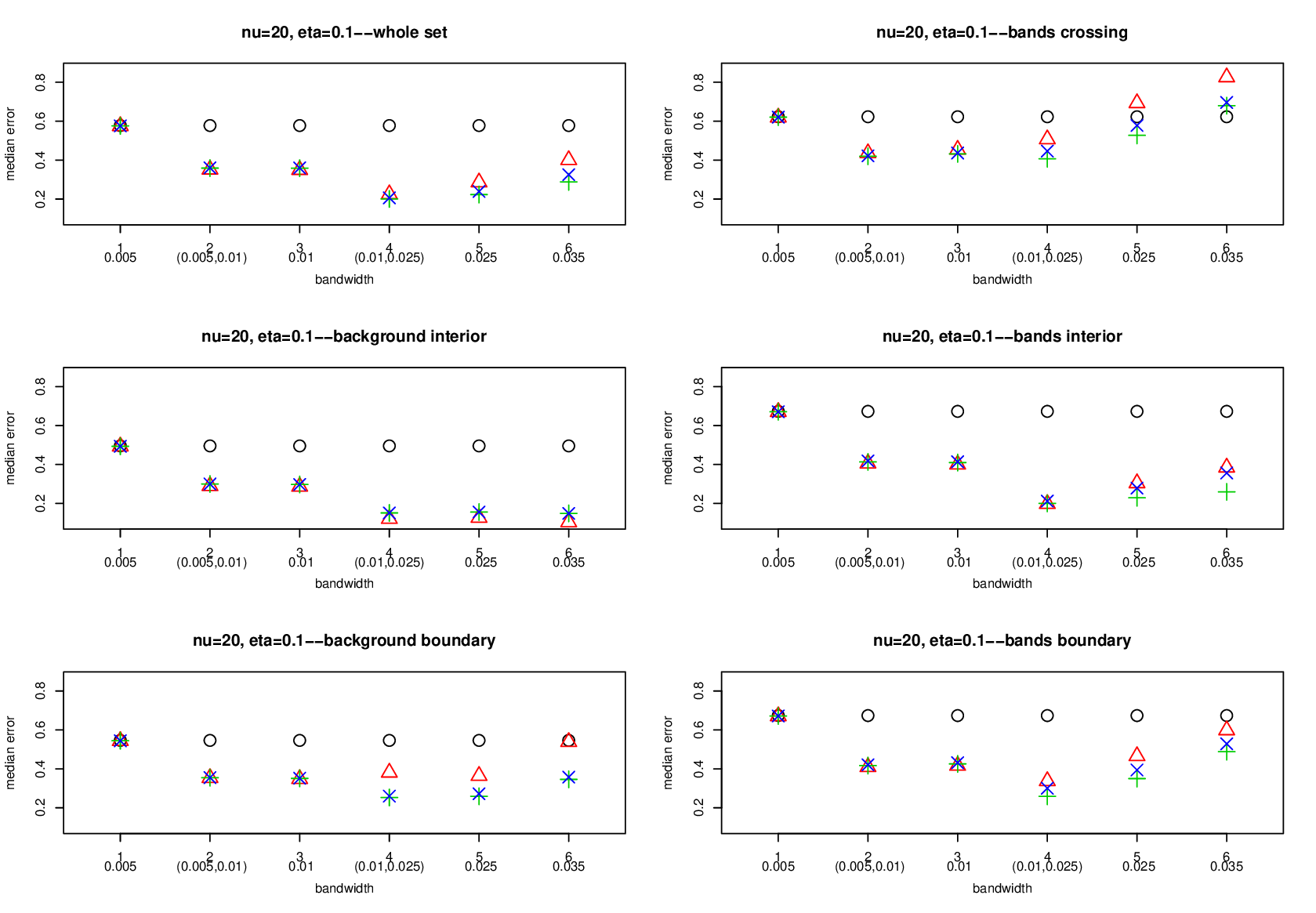}\\
\end{center}
\caption{{\bf Spectral noise} with $\nu = 20$
and $\eta=0.1$. Comparison of median errors over different regions for
observed tensors (black circle) and Euclidean (red triangle), log-Euclidean (green $+$) and
Affine smoothers (blue $\times$).}
\end{figure}

\begin{figure}[th]\label{fig:median_error_spectral_dfeval_20_eta_0.2}
\begin{center}
\includegraphics[height=6.5in,width=5in,angle=270]{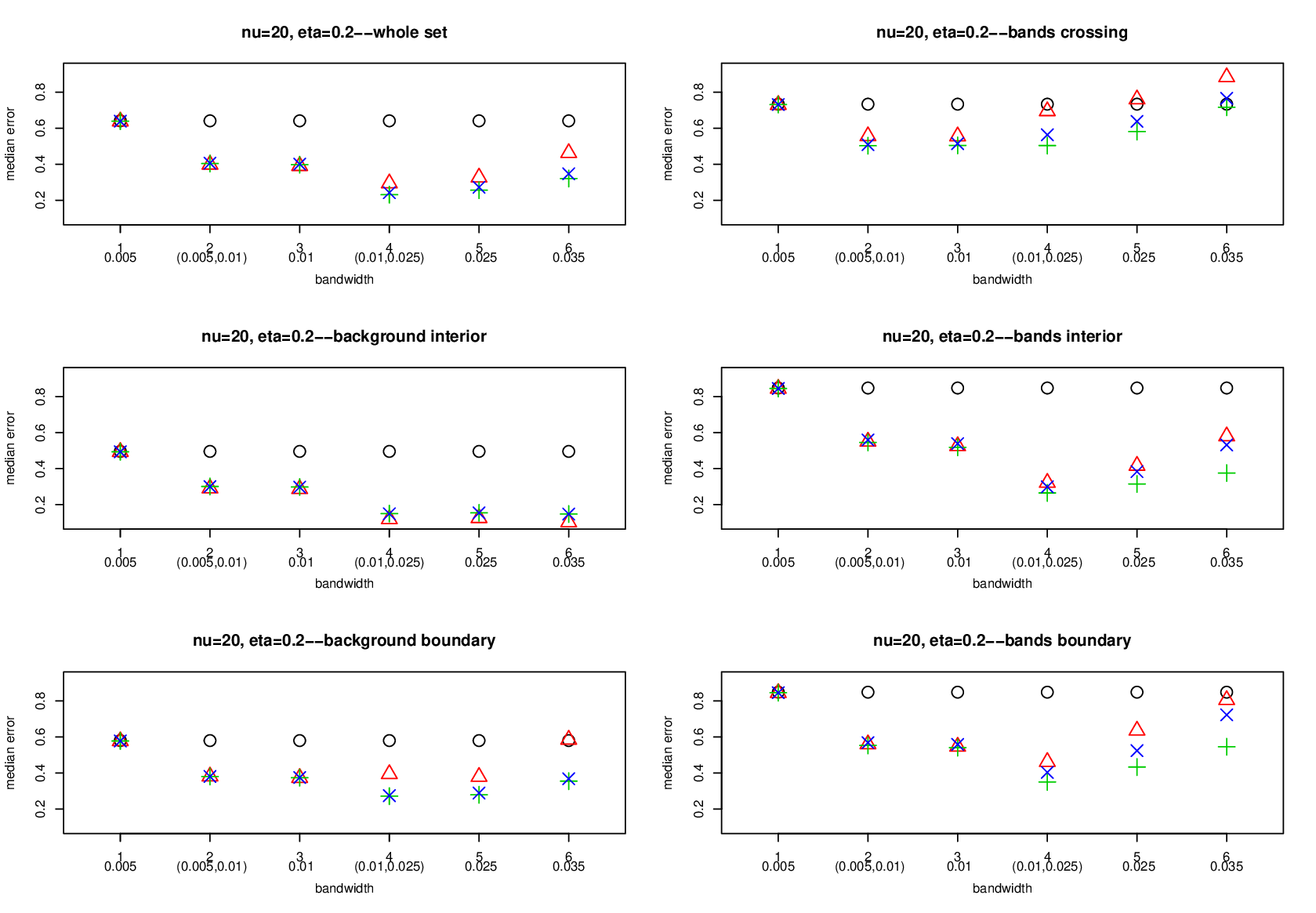}\\
\end{center}
\caption{{\bf Spectral noise} with $\nu = 20$
and $\eta=0.2$. Comparison of median errors over different regions for
observed tensors (black circle) and Euclidean (red triangle), log-Euclidean (green $+$) and
Affine smoothers (blue $\times$).}
\end{figure}

\begin{figure}[th]\label{fig:median_error_spectral_dfeval_20_eta_0.3}
\begin{center}
\includegraphics[height=6.5in,width=5in,angle=270]{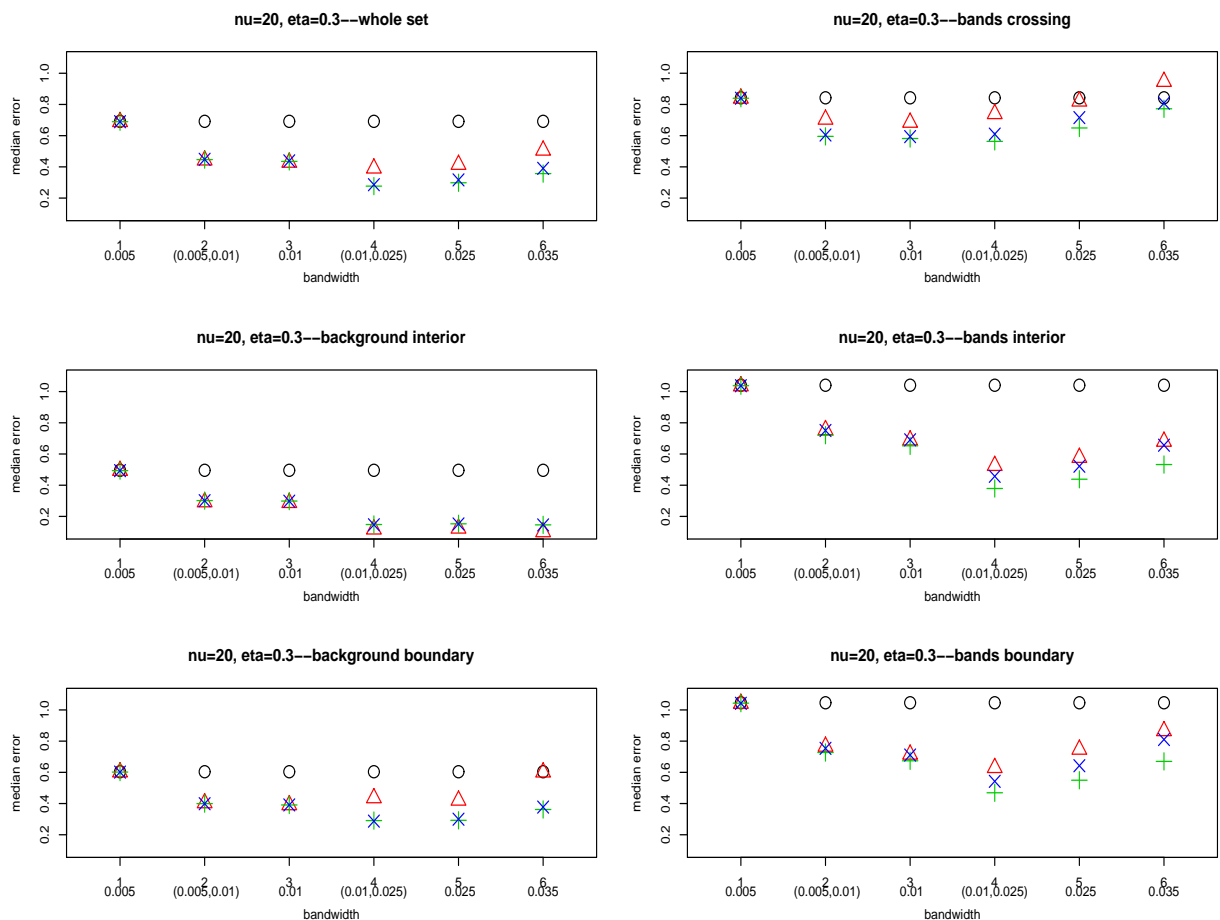}\\
\end{center}
\caption{{\bf Spectral noise} with $\nu = 20$
and $\eta=0.3$. Comparison of median errors over different regions for
observed tensors (black circle) and Euclidean (red triangle), log-Euclidean (green $+$) and
Affine smoothers (blue $\times$).}
\end{figure}

\newpage
\clearpage

\section*{Reference}

\begin{enumerate}

\item
Abramowitz, M. and Stegun, I. (1965). \textit{Handbook of
Mathematical Functions, Ninth Edition}. Dover.

\item
Absil, P.-A., Mahony, R. and Sepulchre, R. (2008).
\textit{Optimization Algorithms on Matrix Manifolds}. Princeton
University Press.

\item
Arsigny, V., Fillard, P., Pennec, X. and Ayache, N. (2005). Fast and
simple computations on tensors with log-Euclidean metrics.
\textit{Technical report \# 5584}. Institut National de Recherche en
Informatique et en Automatique.


\item
Arsigny, V., Fillard, P., Pennec, X. and Ayache, N. (2006).
Log-Euclidean metrics for fast and simple calculus on diffusion
tensors. \textit{Magnetic Resonance in Medicine} {\bf 56}, 411-421.


\item
Basser, P. and Pajevic, S. (2000). Statistical artifacts in
diffusion tensor MRI (DT-MRI) caused by background noise.
\textit{Magnetic Resonance in Medicine} {\bf 44}, 41-50.

\item
Chung, M. K., Lazar, M., Alexender, A. L., Lu, Y. and Davidson, R.
(2003). Probabilistic connectivity measure in diffusion tensor
imaging via anisotropic kernel smoothing.  \textit{Technical
Report}, University of Wisconsin, Madison.

\item
Chung, M. K., Lee, J. E., Alexander, A. L. (2005) : Anisotropic
kernel smoothing in diffusion tensor imaging : theoretical
framework. \textit{Technical report}, University of Wisconsin,
Madison.



\item
Cammoun, L., Castaño-Moraga, C. A., Mu\~{n}oz-Moreno, E.,
Sosa-Cabrera, D., Acar, B., Rodriguez-Florido, M. A., Brun,  A.,
Knutsson,  H. and Thiran,  J. P. (2009). A review of tensors and
tensor signal processing. \textit{Tensors in Image Processing and
Computer Vision : Advances in Pattern Recognition, Part 1}, 1-32.




\item
Fan, J., and Gijbels, I. (1996). {\it Local Polynomial Modelling and
Its Applications}. Chapman \& Hall/CRC.

\item
Ferreira, R., Xavier, J., Costeira, J. P. and Barroso, V. (2006).
Newton method for Riemannian centroid computation in naturally
reductive homogeneous spaces. \textit{Proceedings of ICASSP 2006 -
IEEE International Conference on Acoustics, Speech and Signal
Processing}, Toulouse, France.



\item
Fletcher, P. T. and Joshi, S. (2004). Principal geodesic analysis on
symmetric spaces: Statistics of diffusion tensors. In
\textit{Computer Vision and Mathematical Methods in Medical and
Biomedical Image Analysis, ECCV 2004 Workshops CVAMIA and MMBIA,
Prague, Czech Republic, May 15, 2004}, LNCS vol. 3117, 87–98.
Springer.

\item
Fletcher, P. T. and Joshi, S. (2007). Riemannian geometry for the
statistical analysis of diffusion tensor data. \textit{Signal
Processing} {\bf 87}, 250-262.

\item
F\"{o}rstner, W. and Moonen, B. (1999). A metric for covariance
matrices. In Krumm, F. and Schwarze, V. S., (eds.), \textit{Qua
vadis geodesia...? Festschrift for Erik W. Grafarend on the occasion
of his 60th birthday}, number 1999.6 in Tech. Report of the
Department of Geodesy and Geoinformatics, pages 113–128. Stuttgart
University.

\item
Gudbjartsson, H. and Patz, S. (1995). The Rician Distribution of
Noisy MRI Data. \textit{Magnetic Resonance in Medicine} {\bf 34},
910-914.

\item
Hahn, K. R., Prigarin, S., Heim, S. and Hasan, K. (2006). Random
noise in diffusion tensor imaging, its destructive impact and some
corrections. In \textit{Visualization and Processing of Tensor
Fields}, Eds. Weickert, J. and Hagen, H., 107-117. Springer.

\item
Hahn, K. R., Prigarin, S., Rodenacker, K. and Hasan, K. (2009).
Denoising for diffusion tensor imaging with low signal to noise
ratios: method and monte carlo validation. \textit{International
Journal for Biomathematics and Biostatistics} {\bf 1}(1), 63-81.


\item
Henkelmann, R. M. (1985). Measurement of signal intensities in the
presence of noise in MR images. \textit{Medical Physics} {\bf
12}(2), , 232-233.



\item
Karcher, H. (1977). Riemannian center of mass and mollifier
smoothing. \textit{Communications in Pure and Applied Mathematics}
{\bf 30}, 509-541.

\item
Macorski, A. (1996). Noise in MRI. \textit{Magnetic Resonance in
Medicine} {\bf 36}, 494-497.


\item
Le Bihan, D., Mangin, J.-F., Poupon, C., Clark, A. C., Pappata, S.,
Molko, N. and Chabriat, H. (2001) : Diffusion tensor imaging :
concepts and applications. \textit{Journal of Magnetic Resonance
Imaging} {\bf 13}, 534-546.



\item
Mori, S. (2007). \textit{Introduction to Diffusion Tensor Imaging}.
Elsevier.


\item
Nomizu, K. (1954). Invariant affine connections on homogeneous spaces.
\textit{American Journal of Mathematics} {\bf 76}, 33–65.



\item
Pennec, X., Fillard, P. and Ayache, N. (2006). A Riemannian
framework for tensor computing. \textit{Journal of Computer Vision}
{\bf 66}, 41-66.


\item
Polzehl, J. and Spokoiny, V. (2006). Propagation-separation approach
for local likelihood estimation. \textit{Probability Theory and
Related Fields} {\bf 135}, 335-362.

\item
Polzehl, J. and Tabelow, K. (2008). Structural adaptive smoothing in
diffusion tensor imaging: the R package \texttt{dti}. \textit{WIAS
Technical Report}.

\item
Skovgaard, L. (1984). A Riemannian geometry of the multivariate
normal model. \textit{Scandinavian Journal of Statistics} {\bf 11},
211–223.

\item
Tabelow, K., Polzehl, J., Spokoiny, V. and Voss, H. U. (2008).
Diffusion tensor imaging : structural adaptive smoothing.
\textit{NeuroImage} {\bf 39}, 1763-1773.





\item
Zhu, H. T., Zhang, H. P., Ibrahim, J. G. and Peterson, B. (2007).
Statistical analysis of diffusion tensors in diffusion-weighted
magnetic resonance image data. \textit{Journal of the American
Statistical Association} {\bf 102}, 1081-1110.

\item
Zhu, H. T., Li, Y., Ibrahim, I. G., Shi, X., An, H., Chen, Y., Gao,
W., Lin, W., Rowe, D. B. and Peterson, B. S. (2009). Regression
models for identifying noise sources in magnetic resonance images.
\textit{Journal of the American Statistical Association} {\bf 104},
623-637.
\end{enumerate}

\end{document}